\documentclass[12pt,a4paper,aps,prd,preprint,superscriptaddress,nofootinbib]{revtex4-1}
\usepackage[utf8]{inputenc}
\usepackage{graphicx}
\usepackage{amssymb}
\usepackage{textcomp}
\usepackage{amsmath}
\usepackage{tabularx}
\usepackage{bm}
\usepackage{times}
\usepackage{color}
\usepackage{slashed}
\usepackage{multirow}
\usepackage{verbatim}
\usepackage{cancel}
\usepackage{subfigure}
\usepackage[normalem]{ulem}

\usepackage[colorlinks=true, pdfstartview=FitV, linkcolor=blue, citecolor=blue, urlcolor=blue]{hyperref}
\allowdisplaybreaks[4]

\def\lsim{\mathrel{\raise.3ex\hbox{$<$\kern-.75em\lower1ex\hbox{$\sim$}}}}
\def\gsim{\mathrel{\raise.3ex\hbox{$>$\kern-.75em\lower1ex\hbox{$\sim$}}}}

\newcommand{\bew}{\begin{widetext}}
\newcommand{\enw}{\end{widetext}}
\newcommand{\bee}{\begin{equation}}
\newcommand{\ene}{\end{equation}}
\newcommand{\bea}{\begin{eqnarray}}
\newcommand{\ena}{\end{eqnarray}}
\newcommand{\bes}{\begin{subequations}}
\newcommand{\ens}{\end{subequations}}

\definecolor{deepPurple}{RGB}{150,100,200}
\definecolor{orange}{rgb}{1,0.5,0}

\begin{document}

\title{Searching for long-lived ALPs with a laser-assisted optical dump}

\author{Tong Li}
\email{litong@nankai.edu.cn}
\affiliation{
School of Physics, Nankai University, Tianjin 300071, China
}

\author{Haolong Wang}
\email{2120250183@mail.nankai.edu.cn}
\affiliation{
School of Physics, Nankai University, Tianjin 300071, China
}

\author{Man Yuan}
\email{yuanman@mail.nankai.edu.cn}
\affiliation{
School of Physics, Nankai University, Tianjin 300071, China
}

\begin{abstract}
The feeble interactions of light axion-like particles (ALPs) render them long-lived.
Probing long-lived ALPs therefore demands facilities with a macroscopic decay volume to match their potentially long decay lengths, such as high-intensity beam dump experiments.
An optical dump setup was proposed by utilizing hard photons from the collision of a high-energy electron beam and a high-intensity laser pulse.
In this work, we revisit the probe of long-lived ALPs with MeV $\sim$ GeV mass via a laser-assisted optical dump. We consider the low-energy effective
Lagrangian for ALPs incorporating the ALP-photon and ALP-fermion interactions.
The scope of optical dump searches is extended to both the ALP-photon coupling induced Primakoff process and the Compton-like scattering via the ALP-electron coupling.
We also investigate the correlation
between Primakoff process and Compton scattering, and exhibit the interplay of two ALP couplings in light of optical dump experiment.
\end{abstract}

\maketitle
\tableofcontents

%%%%%%%%%%%%%%%%%%%%%%%%%%%%%%%%
\section{Introduction}
\label{sec:Intro}
%%%%%%%%%%%%%%%%%%%%%%%%%%%%%%%%

The axion-like particle (ALP) is a well-motivated extension of the Standard Model (SM).
It is a CP-odd pseudo-Nambu-Goldstone boson arising from the spontaneous breaking of a global $U(1)$ symmetry. The most well-studied candidate of such CP-odd boson is the QCD axion~\cite{Peccei:1977hh,Peccei:1977ur,Weinberg:1977ma,Wilczek:1977pj} (see a recent review Ref.~\cite{DiLuzio:2020wdo} and references therein). It was suggested to solve the strong CP problem in the Peccei-Quinn (PQ) theory.
The ALP mass ($m_a$) and the symmetry breaking scale (also called decay constant $f_a$) associated with an ALP in many other models can be generally unrelated~\cite{Vysotsky:1978dc,Dimopoulos:1979pp,Tye:1981zy,Zhitnitsky:1980tq,Dine:1981rt,Holdom:1982ex,Kaplan:1985dv,Srednicki:1985xd,Flynn:1987rs,Kamionkowski:1992mf,Berezhiani:2000gh,Hsu:2004mf,Hook:2014cda,Alonso-Alvarez:2018irt,Hook:2019qoh}. The range of ALP mass may vary from sub-micro-eV~\cite{Kim:1979if,Shifman:1979if,Dine:1981rt,Zhitnitsky:1980tq,Turner:1989vc} to TeV scale or even beyond~\cite{Rubakov:1997vp,Fukuda:2015ana,Gherghetta:2016fhp,Dimopoulos:2016lvn,Chiang:2016eav,Gaillard:2018xgk,Gherghetta:2020ofz}.
The nature of ALP depends on the size of various interactions between ALP and the SM matter.
For example, the partial decay width of ALP is essentially given by
\begin{eqnarray}
\Gamma_a\propto
\left\{
  \begin{array}{ll}
    m_a^3/f_a^2\;, & \hbox{for the ALPs coupled to gauge bosons;} \\
    m_a m_f^2/f_a^2\;, & \hbox{for the ALPs coupled to fermion $f$.}
  \end{array}
\right.
\end{eqnarray}
As a result, the ALPs with a small mass at a high-energy scale $f_a$ can manifest as long-lived particles (LLPs). The above scaling implies that long-lived ALPs occupy the corner of parameter space with sufficiently large $f_a$ and small $m_a$ such that the intrinsic width is tiny. They remain detectable only if their boosted decay length matches the macroscopic scale of the experiments. Threshold openings for fermionic channel ($m_a\simeq 2m_f$) also create sharp transitions in lifetime versus mass. This feature makes the boundary between ``prompt'', ``displaced'', and ``invisible'' highly structured rather than smooth~\cite{Bauer:2017ris,FASER:2018eoc,Aloni:2018vki,Cheng:2021kjg,DallaValleGarcia:2023xhh,Li:2025ski}.

The probe of ALPs requires rather different detection strategies and facilities in experiments.
Searches for long-lived ALPs are predominantly conducted via accelerator-based forward search experiments~\cite{Feng:2017uoz,FASER:2018ceo,FASER:2018eoc,FASER:2018bac,FASER:2019aik,FASER:2022hcn,Feng:2022inv,Boiarska:2019vid,Cerci:2021nlb}, or high-intensity beam dump experiments~\cite{Riordan:1987aw,Bjorken:1988as,Bross:1989mp}.
A beam dump experiment places an ultra-dense and thick target with high-atomic-number, typically tungsten (W), lead (Pb), or aluminium (Al), directly in the path of a high-energy primary accelerator beam. The full beam energy is deposited in the dump and generates plentiful secondary SM particles (e.g., mesons, prompt photons, or leptons) from hadronic/electromagnetic cascades inside the dump volume.
The feebly coupled, light, long-lived exotic particles (such as ALPs, dark photons, and etc.) produced in these secondary decays or bremsstrahlung processes can traverse the dense shielding behind the dump without interaction. They then decay or scatter in a low-background, instrumented detector placed several to tens of meters downstream. Unlike limited thin-target setups, beam dumps are engineered to safely absorb the full primary beam intensity. This robust design maximizes the total number of accumulated beam particles and delivers the extreme integrated luminosities essential for probing hidden sector new physics.

Notably, the conventional beam dump experiments are driven by a high-energy accelerator beam of fermionic particle, such as electron~\cite{Riordan:1987aw,Bjorken:1988as,Bross:1989mp,Liu:2017htz,NA64:2020qwq,NA64:2021aiq,Ishikawa:2021qna}, muon~\cite{Sieber:2021fue,NA64:2024klw}, or proton~\cite{CHARM:1985nku,LSND:1997vqj,NOMAD:2001eyx,SHiP:2015vad,Alekhin:2015byh}. The ALP-fermion and ALP-photon couplings induce the ALP production from bremsstrahlung process and Primakoff-like process, respectively, which are both 2 to 3 productions. In particular, the probe of ALP-photon coupling replies on the Primakoff-like process in which a virtual photon is exchanged between the beam particle and the target nuclei~\cite{Dusaev:2020gxi,Liu:2023bby}. An ALP coupled to photons can be emitted from the virtual photon. One usually utilized an Improved Weizs\"{a}cker-Williams (IWW) approximation method to simplify this $2 \to 3$ production to a $2\to 2$ phase space~\cite{Kim:1973he,Bauer:2018onh}. The virtual photon can be viewed as a real photon under this approximation. However, this approximation method results in highly collinear final states and the mass limitation of the produced ALP~\cite{Tsai:1966js,Kim:1973he,Tsai:1973py,Tsai:1986tx,Bauer:2018onh,Liu:2023bby}.

Recently, a new strategy to search for feebly coupled new particles with an optical dump was proposed based on the collision between a high-energy electron beam and a high-intensity laser pulse (shortly named LUXE-NPOD: New Physics at Optical Dump)~\cite{Bai:2021gbm,Soto:2025hnn}. The laser of a high-intensity electromagnetic field promotes the studies of strong-field particle physics and has plenty of applications in high-intensity frontier (see a recent review~\cite{Fedotov:2022ely} and references therein). A high-energy electron beam collides with an intense laser beam of a number of optical photons and radiates a large flux of hard photons. This process is the well-known nonlinear Compton scattering. The observation of nonlinear Compton scattering in the E144 experiment performed at SLAC in 1990s~\cite{Burke:1997ew,Bamber:1999zt} motivated the studies of non-perturbative QED and nonlinear QED~\cite{Hartin:2018egj}. In that LUXE-NPOD proposal, the large flux of hard photons from the proposed Laser Und XFEL Experiment (LUXE)~\cite{Abramowicz:2019gvx,Abramowicz:2021zja} at the European X-Ray Free-Electron Laser (XFEL)~\cite{Altarelli:2006zza} was suggested to strike a dump of target atom tungsten. A bosonic new particle of interaction with photons can be produced through a direct Primakoff process without virtual photon approximation in the calculation~\cite{Bai:2021gbm,Ness:2025klj}. A detector is then placed after the physical dump to detect the diphoton product from the decay of new spin-0 LLP. This method provides us with a good opportunity to actually reexamine the search of long-lived ALP of not merely ALP-photon coupling~\cite{Liu:2022tqn,Liu:2023bby,Pustyntsev:2023rns,Pustyntsev:2024ygw}.

In this work, we investigate the probe of feebly interacting ALPs with photons or fermions via a laser-assisted optical dump. Given ALP-fermion coupling, unlike fermionic beam dump, an ALP can be emitted to final states through a Compton-like scattering between the hard photons and the electrons or nuclei in the target atoms. Thus, one can also explore ALP-fermion coupling via this optical dump experiment. Moreover, this motivates a systematic study of correlations among distinct ALP couplings subject to LLP searches. We consider the low-energy effective Lagrangian for ALP coupled to photon and fermion such as electron. The impact of multiple couplings on ALP decay and lifetime will also be discussed in details. We will first examine the photon-philic or electron-philic case of ALP. The Primakoff process and Compton scattering in optical dump experiment are investigated for the probe of ALP-photon coupling and ALP-electron coupling, respectively. They are both $2 \to 2$ production of long-lived ALP.
We also demonstrate the correlation between Primakoff process and Compton scattering, and exhibit the interplay of two ALP couplings in light of optical dump experiment.

This paper is organized as follows. In Sec.~\ref{sec:ALP}, we review the low-energy effective Lagrangian for ALP coupled to photon or fermion. The details of ALP decay and lifetime will also be discussed.
In Sec.~\ref{sec:Dump}, we consider the laser-assisted production mechanism of hard photons via the nonlinear Compton scattering. The emitted photon spectrum will be presented. We explore the search method of long-lived ALPs at optical dump and show the sensitivity reach for model parameters in Sec.~\ref{sec:Search}. Both Primakoff process and Compton scattering in optical dump experiment will be investigated for the probe of ALP-photon coupling and ALP-electron coupling, respectively. We also consider the correlation
between them and exhibit the interplay of two ALP couplings. Our conclusions are drawn in Sec.~\ref{sec:Con}.

%%%%%%%%%%%%%%%%%%%%%%%%%%%%%%%%%%%%%%%%%
\section{The effective Lagrangian and properties of ALP}
\label{sec:ALP}
%%%%%%%%%%%%%%%%%%%%%%%%%%%%%%%%%%%%%%%%%

In this section, we first review the low-energy effective Lagrangian for an ALP coupled to photons or fermions, and then discuss the resulting implications for ALP properties.

We consider the ALP-photon and ALP-fermion interactions via the following effective dimension-5 Lagrangian below the electroweak scale~\cite{Bauer:2020jbp,Bauer:2021mvw}
\begin{eqnarray}
\mathcal{L}_{\rm ALP}\supset c_{a\gamma}{\alpha \over 4\pi f_a} aF^{\mu\nu}\tilde{F}_{\mu\nu} + c_{af} {\partial_\mu a\over 2f_a} \overline{f} \gamma^\mu \gamma_5 f\;,
\end{eqnarray}
where $c_{a\gamma}$ and $c_{af}$ are two dimensionless parameters, $\alpha$ is the fine-structure constant, $F_{\mu\nu}$ and $\tilde{F}_{\mu\nu}$ are the electromagnetic field strength tensor and its Hodge dual tensor, respectively.
The above ALP-photon coupling is a model-dependent ultraviolet (UV) parameter and is related to the electromagnetic anomaly of the global $U(1)$ symmetry.
In this work, the ALP-photon coupling is simply taken as an independent parameter in an effective framework. A dimensional UV ALP-photon coupling is defined as $g_{a\gamma}\equiv c_{a\gamma}\alpha/(\pi f_a)$. The ALP-fermion interacting term can be rewritten up to a total derivative and equation of motion
\begin{eqnarray}
c_{af} {\partial_\mu a\over 2f_a} \overline{f} \gamma^\mu \gamma_5 f \to -i {c_{af}m_f\over f_a} a \overline{f} \gamma_5 f\;.
\end{eqnarray}
One usually defines another dimensionless ALP-fermion coupling $g_{af}\equiv c_{af}m_f/f_a$. In this work, we restrict us to only ALP-electron interaction. Flavor non-universality of fermions is expected in the most general effective Lagrangian for fermionic ALP interactions~\cite{Georgi:1986df,Bauer:2021mvw} because non-universal PQ charge matrices can be embedded to the mass eigenstate basis. In particular, a class of models relate the $U(1)_{\rm PQ}$ symmetry to a global $U(1)$ flavor symmetry~\cite{Berezhiani:1989fp,Ema:2016ops,Calibbi:2016hwq,Bjorkeroth:2018ipq,delaVega:2021ugs}. The electron-philic ALP would emerge by assigning specific PQ charge for SM leptons.
As a phenomenological study, we remain agnostic about the origin of the ALP-electron coupling.

\begin{figure}[htp]
\centering
\raggedright
\includegraphics[width=0.44\textwidth]{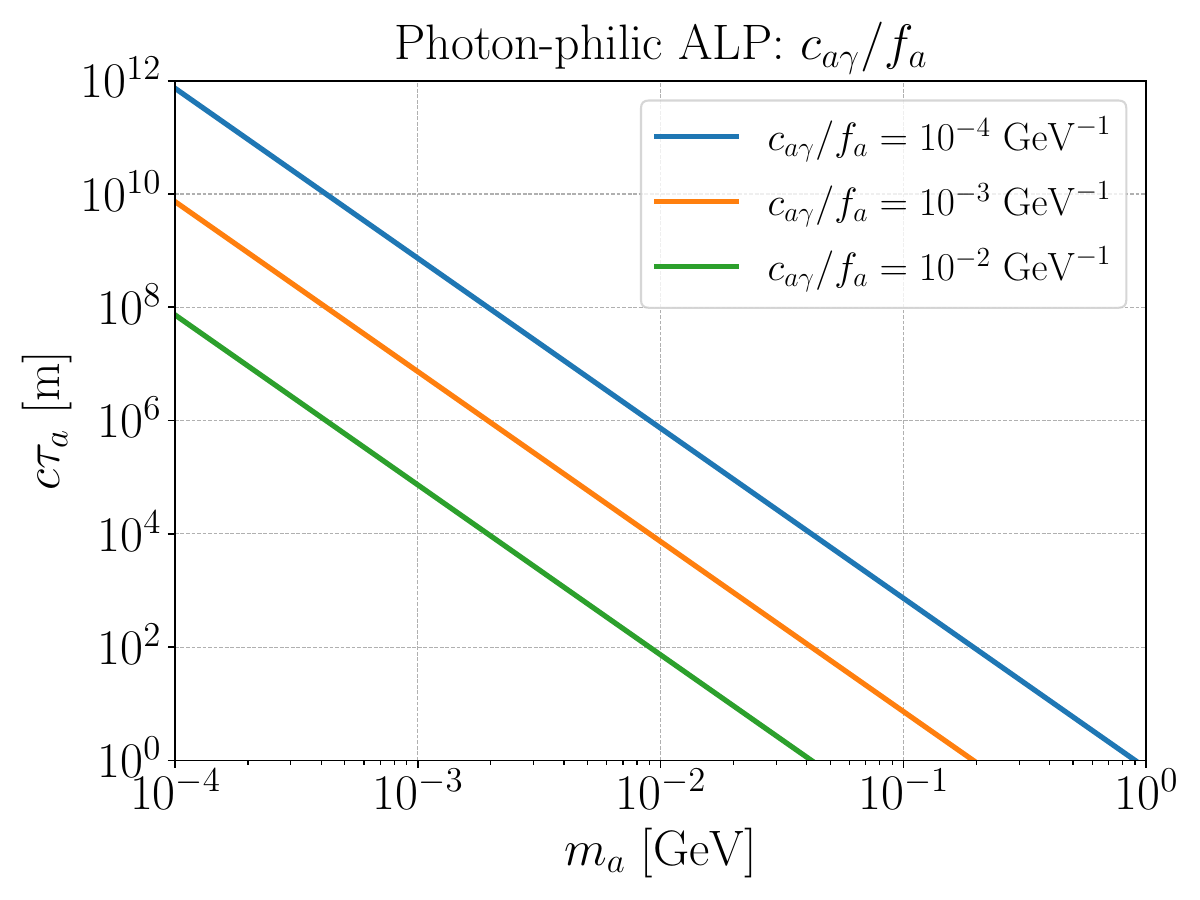}\\
%\end{flushleft}
%\vspace{0.4em}
\includegraphics[width=0.44\textwidth]{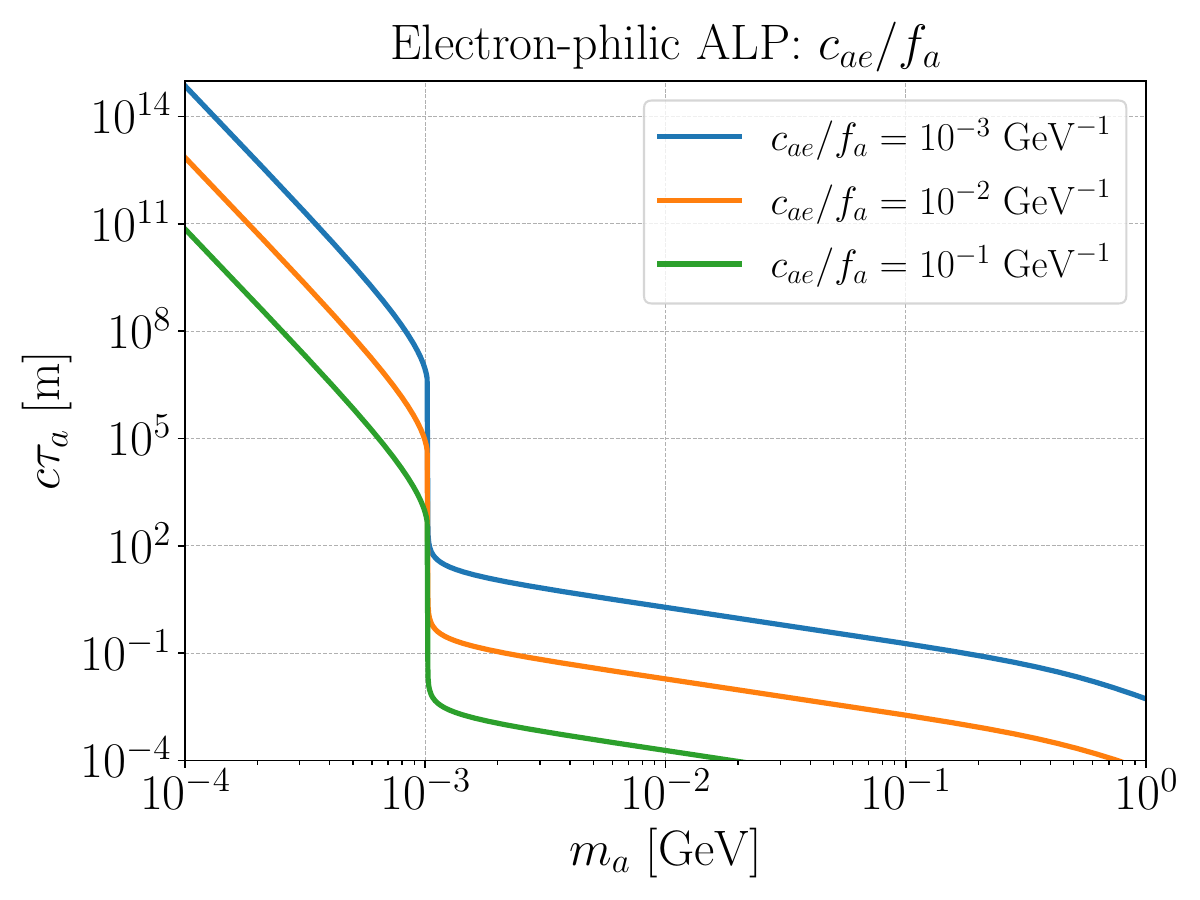}
%\hfill
\includegraphics[width=0.44\textwidth]{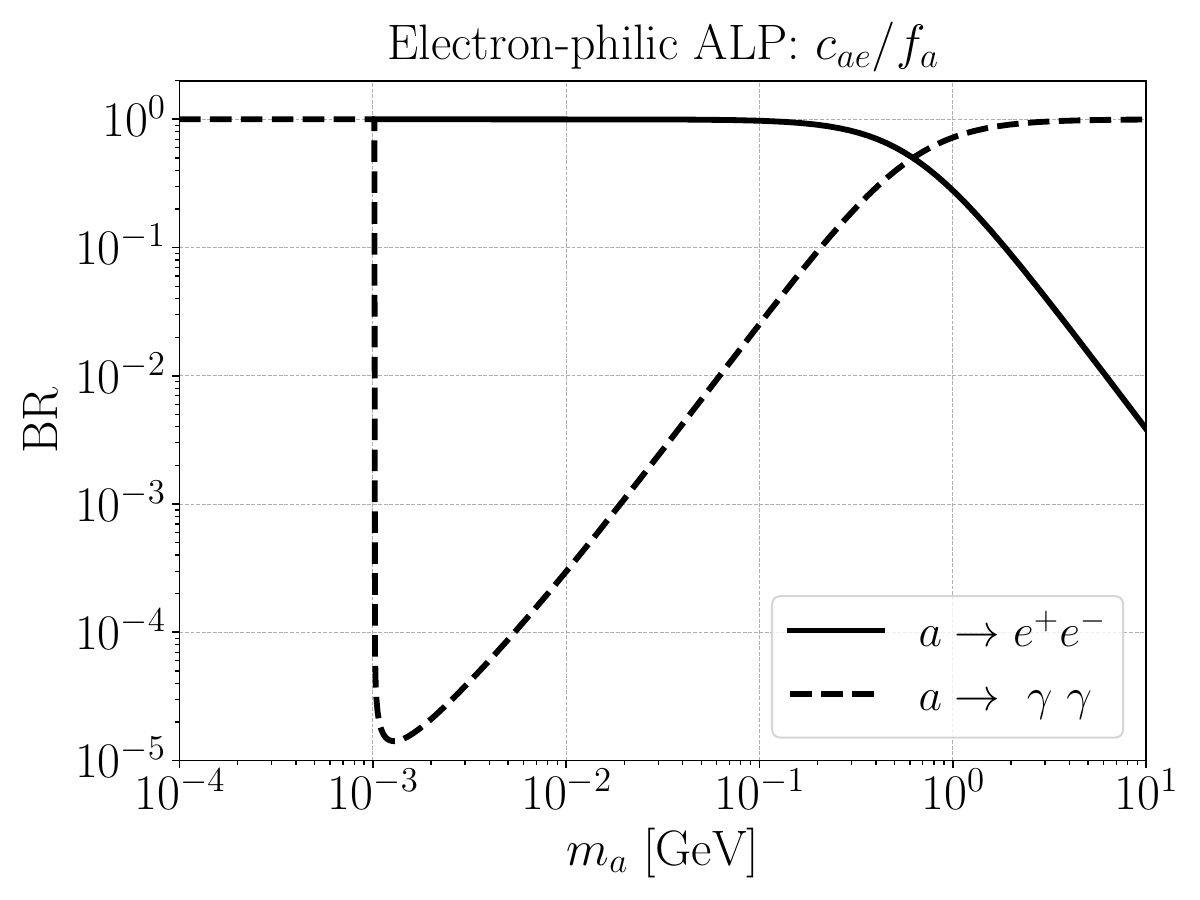}\\
%\vspace{0.4em}
\includegraphics[width=0.44\textwidth]{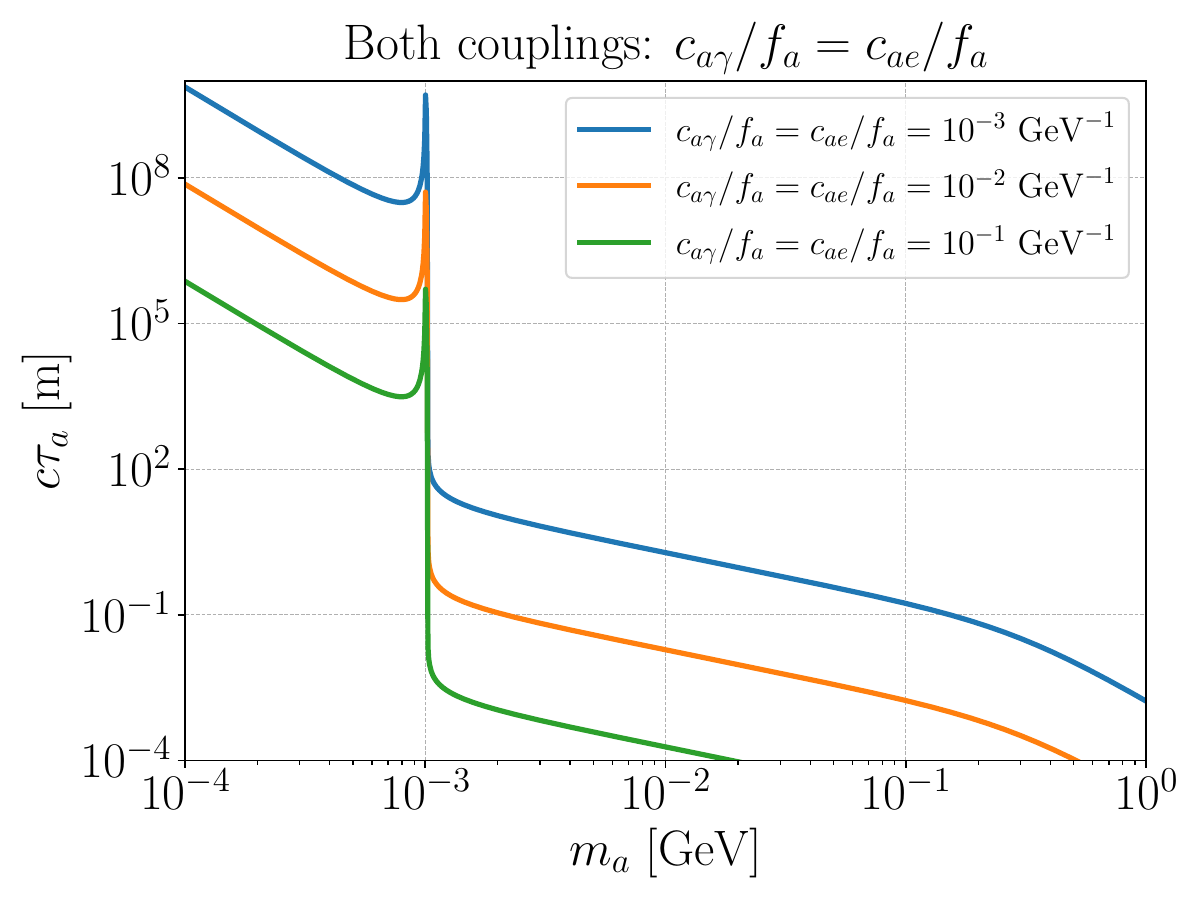}
%\hfill
\includegraphics[width=0.44\textwidth]{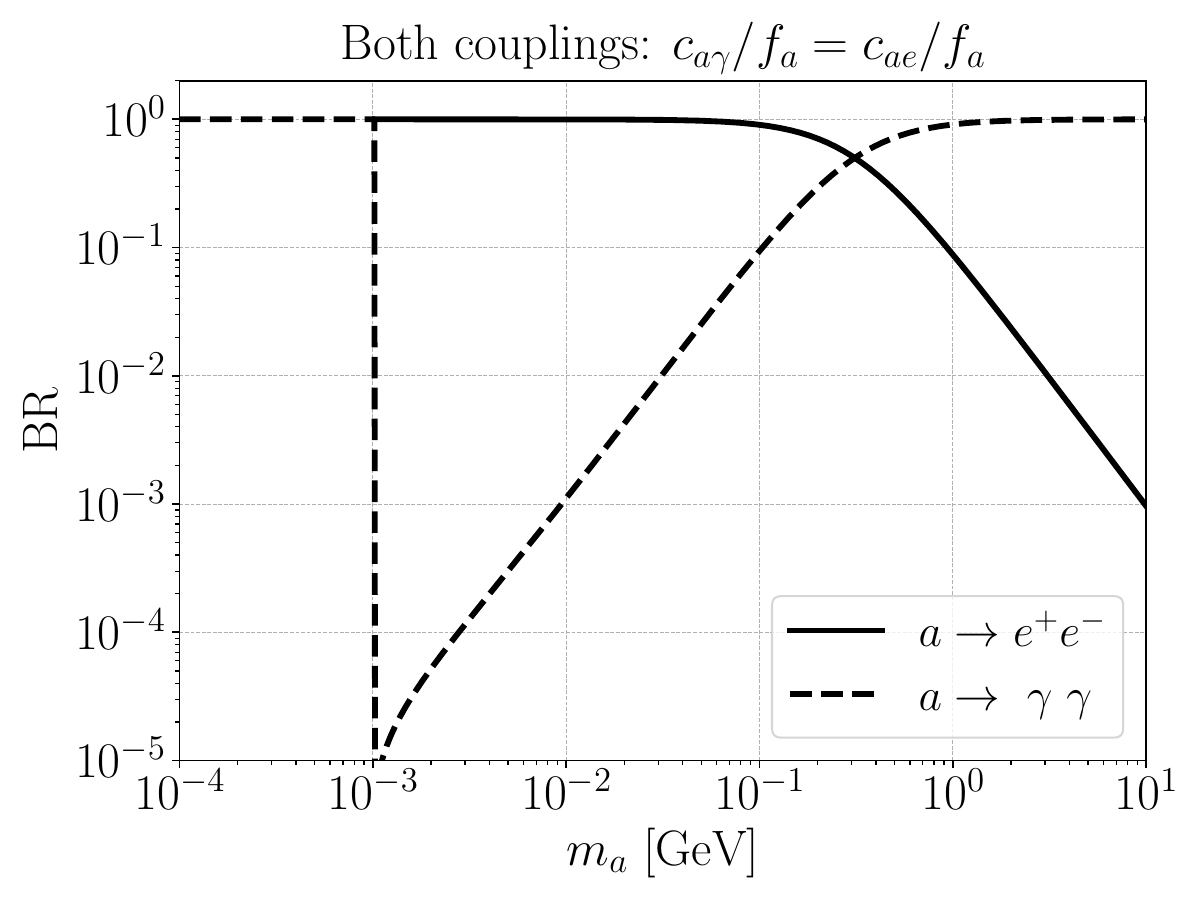}
\caption{
Proper decay length $c\tau_a$ (left panels) and decay branching ratios (right panels) as functions of the ALP mass $m_a$.
The first row shows only $c\tau_a$ for the photon-philic case with only $c_{a\gamma}/f_a$, the second row corresponds to the electron-philic case with only $c_{ae}/f_a$, and the third row is for the case with both couplings satisfying $c_{a\gamma}/f_a=c_{ae}/f_a$.
In the branching-ratio panels, the solid and dashed curves denote $a\to e^+e^-$ and $a\to\gamma\gamma$, respectively.
}
\label{fig:ctau_br}
\end{figure}

Since light ALPs can only be produced in beam dump experiments if their mass lies below GeV scale, we focus our analysis on the $a\to \gamma\gamma$ and $a\to e^+e^-$ decay channels.
For photon-philic or electron-philic scenario, the ALP decay width is respectively given by
\begin{eqnarray}
\Gamma(a\to\gamma\gamma)&=&\frac{m_a^3 \alpha ^2 }{64 \pi ^3 f_a^2} c_{a\gamma}^2={m_a^3\over 64\pi}g_{a\gamma}^2\;,\\
\Gamma(a\to e^+ e^-)&=&{m_a m_e^2\over 8\pi f_a^2}c_{ae}^2 \sqrt{1-{4m_e^2\over m_a^2}}={m_a\over 8\pi}g_{ae}^2 \sqrt{1-{4m_e^2\over m_a^2}}~~~{\rm for~}m_a>2m_e\;.
\end{eqnarray}
Notably, the ALP-electron coupling induces an additional loop correction to the $a\to \gamma\gamma$ decay width. When both couplings exist, the $a\to\gamma\gamma$ decay width becomes~\cite{Bauer:2021mvw}
\begin{eqnarray}
\Gamma(a\to\gamma\gamma)=\frac{m_a^3 \alpha ^2 }{64 \pi ^3 f_a^2}\left|c_{a\gamma}+c_{ae} B\left( \frac{4m_{e}^2}{m_a^2}\right)\right|^2=\frac{m_a^3 \alpha ^2 }{64 \pi ^3 f_a^2} \left| c_{a\gamma}^{\rm eff}\right|^2,
\label{eq:diphoton}
\end{eqnarray}
where the effective coupling is $c_{a\gamma}^{\rm eff}\alpha/(\pi f_a)\equiv g_{a\gamma}^{\rm eff}$ as the combination of UV ALP-photon coupling $c_{a\gamma}$ and $c_{ae}$ induced loop correction, and the loop function $B$ reads
\begin{align}
    B(x)= 1-x f(x)^2,~~ f(x)=\begin{cases}
 \arcsin\left(\frac{1}{\sqrt{x }}\right) & x\geq 1 \,,\\
 \frac{\pi }{2}+\frac{i}{2} {\rm ln} \left(\frac{1+\sqrt{1-x}}{1-\sqrt{1-x}}\right) & x <1 \,.
\end{cases}
\label{eq:B}
\end{align}
The electron-loop term in Eq.~(\ref{eq:diphoton}) is unavoidable for $a\to \gamma\gamma$ decay width if ALP-electron coupling exists.
The ALP-photon coupling can also induce an additional loop-level contribution to the ALP-electron coupling. However, according to the derivation in Ref.~\cite{Bauer:2017ris}, the ALP-photon coupling induced part in effective ALP-electron coupling is proportional to $\alpha^2 c_{a\gamma}/(16\pi^2)$. Compared with $c_{ae}$ of the same order as $c_{a\gamma}$, the ALP-photon coupling induced contribution can be neglected as shown in other literature (for instance Ref.~\cite{Bauer:2021mvw}). We thus ignore this loop-level contribution in our calculation.
In Fig.~\ref{fig:ctau_br}, we show the ALP decay lengths (left panels) and the ALP decay branching ratios (right panels) as a function of $m_a$ for photon-philic case (top, with a branching ratio of unity into $\gamma\gamma$), electron-philic case (middle) and the case in the presence of both couplings (bottom).
For illustration, we choose representative benchmark values $c_{a\gamma}/f_a=\{10^{-4},$ $10^{-3},10^{-2}\}~{\rm GeV}^{-1}$ in the photon-philic case, $c_{ae}/f_a=\{10^{-3},10^{-2},10^{-1}\}~{\rm GeV}^{-1}$ in the electron-philic case, and $c_{a\gamma}/f_a=c_{ae}/f_a=\{10^{-3},10^{-2},10^{-1}\}~{\rm GeV}^{-1}$ in the two-coupling case.
In the branching-ratio panels, the solid and dashed curves denote the $a\to e^+e^-$ and $a\to\gamma\gamma$ decay modes, respectively.
The lifetime panels show that $c\tau_a$ decreases with increasing $m_a$ and coupling strength.
In the electron-philic and two-coupling cases, the opening of the tree-level $a\to e^+e^-$ channel at $m_a=2m_e\approx 1$ MeV significantly shortens the ALP lifetime.
The branching-ratio panels show that the diphoton mode dominates below the electron threshold, while the $e^+e^-$ channel becomes important above threshold.
For $m_a$ greater than about 0.1 GeV, the diphoton branching ratio increases again due to the faster mass scaling of the diphoton decay width.
These features determine both the ALP decay probability in the downstream detector and the relevant visible final states.
Note that the inclusion of muon and tau leptons will modify the ALP decay features when kinematically allowed. Nevertheless, they involve more free coupling parameters and only change the edge of the our sensitivity reach for $m_a\gtrsim 0.2$ GeV and $m_a\gtrsim 3.5$ GeV. We will thus not take into account their contribution.

%%%%%%%%%%%%%%%%%%%%%%%%%%%%%%%%%%%%%%%%%%%%%%%%
\section{Laser-assisted photon source and optical dump }
\label{sec:Dump}
%%%%%%%%%%%%%%%%%%%%%%%%%%%%%%%%%%%%%%%%%%%%%%%%%

In this section, we discuss the laser-assisted collision and the consequent high-energy photon spectrum for the optical dump strategy proposed for the LUXE experiment in Ref.~\cite{Bai:2021gbm}.
In this optical dump proposal, the high flux of GeV photons generated in an electron-laser collision is used as the beam source for the LLP search. For this setup, an electron bunch collides with an optical laser pulse and produces a forward photon beam with GeV-scale energies. The emitted photons are then directed onto a physical dump, where interactions with the target material can produce new weakly coupled particles.

The production of the hard photon beam is described within the framework of strong-field QED. In the external background of high-intensity laser, the incoming electron undergoes continuing interactions with plentiful laser photons and is treated as a Volkov state~\cite{Wolkow1935}, denoted by $e_V^-$. The dominant emission process is through nonlinear Compton scattering~\cite{Burke:1997ew,Bamber:1999zt}
\begin{eqnarray}
e_V^- (+ n\gamma_{\rm Laser})\to e_V^- +\gamma \;.
\end{eqnarray}
In principle, the emitted photons can initiate nonlinear Breit-Wheeler pair production~\cite{Reiss:1962nhe} inside the laser field.
For the LUXE benchmark parameters, this process occurs on a much longer time scale than photon emission and the laser-pulse duration~\cite{Bai:2021gbm,Abramowicz:2021zja}.
Therefore, the laser behaves as an effective thick medium for the electrons, while most of the emitted photons stream freely through it.
This is the key feature of the optical dump setup: an intense hard-photon beam is formed before the downstream physical dump.
Note that the collision of an electron beam and intense laser pulse can also directly produce a new massive dark particle beyond the SM~\cite{Dillon:2018ypt,King:2018qbq,Dillon:2018ouq,Beyer:2021mzq,Ma:2024ywm,Bai:2021gbm,Ness:2025klj} through the interaction with electron, e.g. $e_V^-\to e_V^- + a$ via ALP-electron coupling. However, the mass of produced dark particle is restricted to be less than $2m_e\sim 1$ MeV~\cite{Ma:2024ywm}. As a result, the lifetime of the dark particle is sufficiently long so that it typically decays well beyond the detector located several meters downstream of the dump. We will not consider this ``primary'' production mode below.

For the later calculation, we distinguish two interaction areas in the optical dump setup: the photon source area and the ALP production area. The electron-laser collision area provides the primary photon beam and fixes the spectrum $dN_\gamma/dE_\gamma$, normalized per incident electron.
For electron beam energy $E_e=16.5~{\rm GeV}$, we obtain the primary photon spectrum using Monte Carlo package Ptarmigan~\cite{ptarmigan.2023,Blackburn2026ptarmigan}, which implements the strong-field QED simulation of nonlinear Compton photon emission at LUXE experiment.
The input parameters of Ptarmigan are based on the LUXE electron-beam and laser benchmark configurations~\cite{Abramowicz:2021zja,Bai:2021gbm}.
The electron beam energy is set to $E_e=16.5~{\rm GeV}$, and each electron bunch contains $N_e=1.5\times10^9$ physical electrons.
The remaining electron-beam parameters are taken from the LUXE baseline setup, including a transverse beam size of $5~\mu{\rm m}$ and a bunch length of $24~\mu{\rm m}$~\cite{Abramowicz:2021zja}.
Two operation phases were suggested for the LUXE experiment~\cite{Abramowicz:2021zja,Bai:2021gbm}: phase-0 (phase-1) refers to the initial 40 TW (upgraded 350 TW) LUXE laser setup.
The laser intensity parameter is defined as $\xi \equiv eE/(m_e\omega_L)$ with $E$ being the peak electric field and $\omega_L$ the laser photon energy.
The phase-0 (phase-1) laser input uses intensity parameter $\xi=3.2~(3.4)$, laser wavelength $\lambda=0.8~\mu{\rm m}$, pulse duration $\tau=25~(120)~{\rm fs}$, transverse spot size $w_0=6.5~(10)~\mu{\rm m}$.
The laser polarization is circular in both cases.

\begin{figure}[t]
\centering
\includegraphics[width=0.48\textwidth]{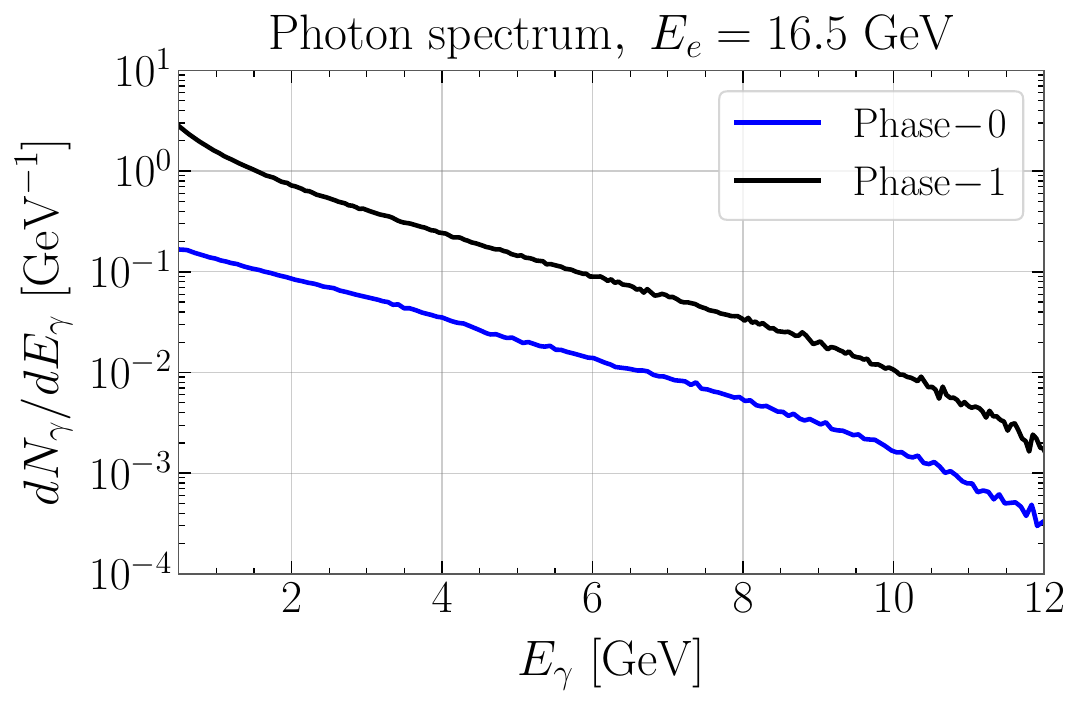}
\includegraphics[width=0.48\textwidth]{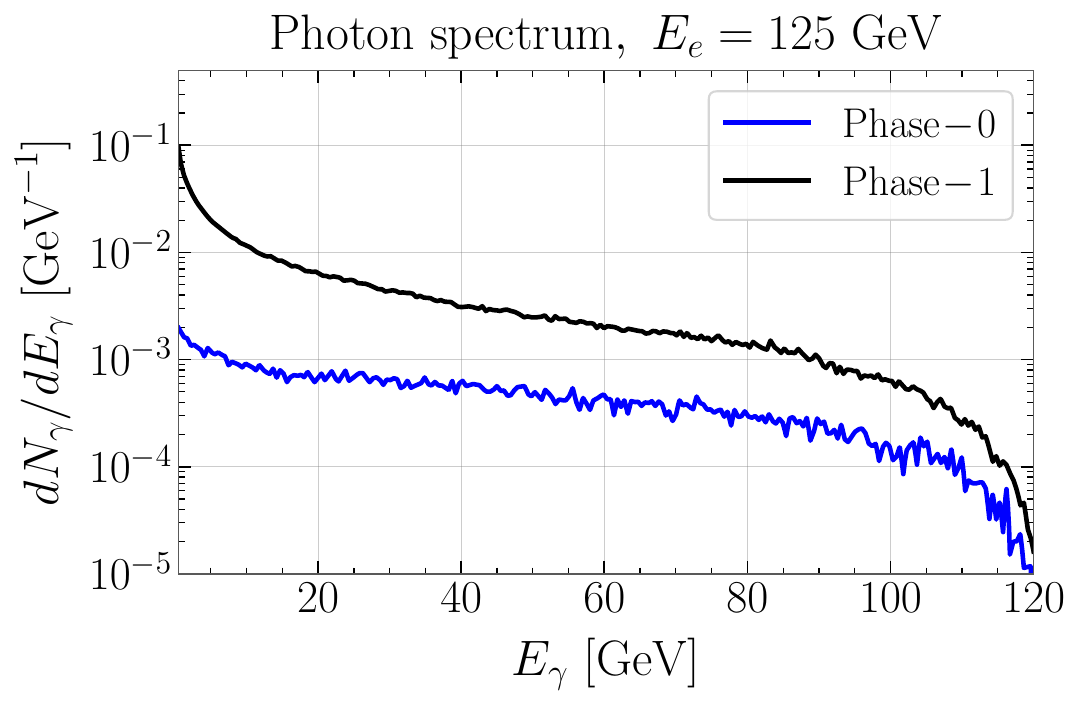}
\caption{
Photon energy spectra $dN_\gamma/dE_\gamma$ generated in the electron-laser interaction area, normalized per incident electron. The left panel shows the $E_e=16.5~{\rm GeV}$ LUXE benchmark used in Ref.~\cite{Bai:2021gbm}. The right panel shows the $E_e=125~{\rm GeV}$ benchmark adopted in this work as a representative high-energy scenario. In each panel, the blue and black curves denote the phase-0 and phase-1 laser configurations, respectively.
}
\label{fig:photon-spectrum}
\end{figure}

The resulting spectra in phase-0 and phase-1 are shown in the left panel of Fig.~\ref{fig:photon-spectrum}, which agrees with the expected LUXE benchmark distribution~\cite{Bai:2021gbm}.
Besides, high-energy electron beams are being developed to enable studies in high-precision and high-energy elementary particle physics~\cite{NA64:2020qwq,NA64:2021aiq,ILC:2013jhg,FCC:2018evy,CEPCStudyGroup:2018ghi}. We also consider a higher energy benchmark with $E_e=125~{\rm GeV}$ as a representative future scenario~\cite{Irles:2023udu,Schulthess:2025tct}.
The electron-beam setups are taken from the baseline parameters for the International Linear Collider (ILC)~\cite{Adolphsen:2013kya,Irles:2023udu,Schulthess:2025tct}: bunch electrons $N_e=2\times 10^{10}$, transverse beam size 75 nm given as the geometric mean of horizontal and vertical beam sizes, and bunch length $300~\mu{\rm m}$. The laser properties for phase-0 and phase-1 are assumed to be the same as LUXE benchmarks.
The resulting phase-0 and phase-1 spectra for this high-energy electron beam are shown in the right panel of Fig.~\ref{fig:photon-spectrum}.
Both spectra decrease with the photon energy, but they retain a sizable photon flux over the energy range of GeV-scale (tens-of-GeV) relevant for ALP production with $E_e=16.5~(125)$ GeV. For a fixed electron-beam energy, the phase-1 spectrum is significantly larger than the phase-0 spectrum across the full range. This reflects the enhanced photon yield expected in the upgraded laser configuration.
This high-energy photon spectrum is the common input for all subsequent event-rate calculations in the physical dump.

The downstream physical dump is the area of ALP production, where the incoming photons scatter off tungsten nucleus or atomic electrons and may be converted into ALPs. In the original LUXE-NPOD study, this secondary production was mainly considered through the Primakoff process between photons and nucleus $N$, $\gamma+N\to a+N$, for spin-0 particles coupled to photons~\cite{Bai:2021gbm}. In this work, we also consider Compton-like scattering which allows the optical dump setup to probe ALP couplings to fermion.

%%%%%%%%%%%%%%%%%%%%%%%%%%%%%%%%%%%%%%%%%%%
\section{Searching for long-lived ALPs at optical dump}
\label{sec:Search}
%%%%%%%%%%%%%%%%%%%%%%%%%%%%%%%%%%%%%%%%%%%

After the hard photon is generated in the electron-laser interaction area, it is used as the incident beam for ALP production in the downstream dump. In this section, we describe the production and decay probability of long-lived ALPs in the optical dump setup. We then calculate the visible signal yield and present the projected sensitivity to ALP couplings.

We now evaluate the total production rate of ALPs within the experimental volume. The total detectable ALP yield, $N_a$, determines the sensitivity reach of the proposed optical dump experiment. By convolving the primary photon spectrum with the target-specific production cross section, accounting for the decay probability within the detector's volume and the angular acceptance and efficiency of the detector, the visible signal yield for a given production channel $i$ can be written as
\begin{eqnarray}
N_a^i &\approx& \mathcal{L}_{\text{eff}} \int dE_\gamma~\frac{dN_\gamma}{dE_\gamma} \int_{t_{\rm min}}^{t_{\rm max}} dt~\frac{d\sigma^i(E_\gamma)}{dt}~\mathcal{P}_{\rm decay} ~\mathcal{A}\;,
\label{eqn:Na}
\end{eqnarray}
where $i=$ ``Prim'' and ``Comp'' denote the Primakoff process and Compton scattering, respectively.
Each term in Eq.~(\ref{eqn:Na}) is defined as follows~\cite{Bai:2021gbm}
\begin{itemize}
\item The effective luminosity $\mathcal{L}_{\rm eff}$ incorporates the electron bunch population $N_e=1.5 \times 10^9$ electrons with  a fixed energy of $E_e=16.5$ GeV, $N_{\rm BX}=1.0 \times 10^7$ laser-pulse and electron-bunch collisions per year and other target properties~\cite{Tsai:1973py}
\begin{eqnarray}
\mathcal{L}_{\text{eff}} &=& N_e  N_{\rm BX} \frac{9\rho_W X_0}{7 A_W m_0}\;,
\label{eq:Lum}
\end{eqnarray}
where $\rho_W = 19.3 ~{\rm g}/{\rm cm}^3$ is tungsten density, ${9\over 7}X_0 \approx 0.45 ~{\rm cm}$ is the radiation length of photons losing most of their energy, $A_W = 184$ and $m_0 = 1.66 \times 10^{-24} \text{ g}$ ($\sim 930 \text{ MeV}$) represent the mass number of tungsten and the nucleon mass, respectively. The resulting effective luminosity is approximately estimated to be $\mathcal{L}_{\rm eff} \approx 0.426 ~{\rm fb}^{-1}$. For an ILC-like high-energy electron beam, we instead have $E_e=125$ GeV, $N_e=2\times 10^{10}$ and $N_{\rm BX}=6.6\times 10^{10}$~\cite{Irles:2023udu}. The consequent effective luminosity becomes $\mathcal{L}_{\rm eff} \approx  3.753 \times10^4~{\rm fb}^{-1}$. For Eqs.~(\ref{eqn:Na}) and (\ref{eq:Lum}), we use the formulas in LUXE-NPOD proposal~\cite{Bai:2021gbm} which follows the Eq.~(17) of Ref.~\cite{Chen:2017awl} for muon beam and the Eq.~(45) of Ref.~\cite{Berlin:2018pwi} for secondary photons. The involvement of effective luminosity agrees with the signal rate in Ref.~\cite{Chen:2017awl} with a constant energy loss and in Ref.~\cite{Cesarotti:2022ttv} under the assumption of zero radiative losses of the beam particle in the material.
\item The differential energy spectrum of primary photons $dN_\gamma/dE_\gamma$, constituting a foundational input for the ALP yield calculation, represents the number of high-energy photons generated per unit energy interval within the laser-electron collision. These photons are created via nonlinear Compton scattering-a cornerstone strong-field QED process in which a relativistic electron coherently absorbs $n \geq 1$ optical laser photons and emits a single high-energy photon $e^- + n\gamma_{\rm Laser} \to e^- + \gamma$. We adopt the photon energy spectrum obtained by Ptarmigan package in Sec.~\ref{sec:Dump}.
\item The cross section $d\sigma^i/dt$ describes the conversion of an incoming photon into an ALP through the production channel $i$. For Primakoff production with $i=$ ``Prim'', the photon scatters off the electromagnetic field of the tungsten nucleus and the rate is controlled by the effective ALP-photon coupling $g_{a\gamma}^{\rm eff}$. For Compton-like production with $i=$ ``Comp'', the incoming photon scatters off a target fermion. In this work, we consider atomic electrons outside the nucleus for probing the ALP-electron coupling $g_{ae}$. The explicit form of the differential cross section will be given below.
\item The decay probability factor is given by
\begin{eqnarray}
\mathcal{P}_{\rm decay} = \Big(e^{-L_D / L_a} - e^{-(L_V + L_D) / L_a}\Big)\times {\rm BR}(a\to X)\;,
\end{eqnarray}
where $L_D = 1$ m and $L_V = 2.5$ m are the characteristic length of the tungsten dump and the distance from dump end to downstream detector, respectively. $L_a = c\, \tau_a\, p_a / m_a$ denotes the decay length of the ALP with $\tau_a$ being the ALP lifetime and $p_a = \sqrt{E_a^2 - m_a^2}$ its momentum. ${\rm BR}(a\to X)$ denotes the branching ratio of ALP decay to final state $X=\gamma\gamma$ or $e^+e^-$. This factor requires the ALP to escape the dump and decay inside the downstream decay volume. The LUXE-NPOD proposal suggested to place a magnet placed at the back of the dump aiming to bend away the charged SM particles produced in the dump and reduce the generated SM backgrounds. Nevertheless, as discussed in the layout optimization for LUXE-NPOD~\cite{Soto:2025hnn}, the charged SM component is subdominant in backgrounds and thus it does not change the overall picture of the background, even if an additional magnet is not considered. Suppose the magnet is really placed right after the bump, to account for the effect of the magnetic field for $e^+e^-$ from ALP decay, we can modify the decay probability factor to explore the ALP preferring to decay after the magnet
\begin{eqnarray}
\mathcal{P}_{\rm decay} = \Big(e^{-(L_D+L_M) / L_a} - e^{-(L_V + L_D) / L_a}\Big)\times {\rm BR}(a\to e^+e^-)\;,
\end{eqnarray}
where $0<L_M<L_V$ denotes the length of shielding magnet and $L_M=0.5$ m is assumed to estimate the $e^+e^-$ signal rate.
\item $\mathcal{A}$ denotes the angular acceptance and efficiency of the detector. We assume a photon energy detection threshold of $E_{\gamma} = 0.5$ GeV and take the angular acceptance to be $\mathcal{A} = 1$. This is justified because the typical signals of $\mathcal{O}(1-10)$ GeV photon energy for ALPs with mass of $\mathcal{O}(1-10^5)$ MeV produce angular distributions largely compatible with the LUXE-NPOD detector geometry. Any resulting geometric corrections are at most of $\mathcal{O}(1)$ and do not significantly affect our results.
\end{itemize}

Next, we discuss the individual production channels and show the relevant sensitivity reach of ALP couplings via the optical dump.

%%%%%%%%%%%%%%%%%%%%%%%%%%%
\subsection{Primakoff production}
%%%%%%%%%%%%%%%%%%%%%%%%%%%

The hard photon beam collides the physical dump and can produce light ALPs through two different processes: Primakoff production or Compton scattering. We will first investigate these two production channels induced by individual ALP coupling, and analyze their possible correlation in next subsection.

\begin{figure}
\centering
\includegraphics[width=0.60\textwidth]{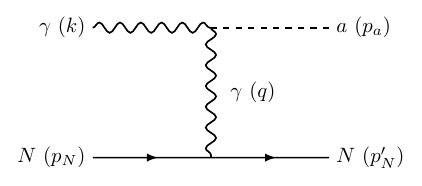}
\caption{
The Feynman diagram for ALP production via photon scattering off a nucleus, $\gamma(k) + N(p_N) \to a(p_a) + N(p_N')$, induced by the ALP-photon coupling $g_{a\gamma}$.
}
\label{fig:feynman_gag}
\end{figure}

For the photon-philic ALP only coupled to photons, the relevant production channel is Primakoff process from the collision between photons and dump nucleus $N$
\begin{eqnarray}
\gamma(k)+N(p_N)\to a(p_a)+N(p'_N)\;, \qquad q=k-p_a=p'_N-p_N\;,
\end{eqnarray}
with the Mandelstam variables $t=q^2$ and $s=(k+p_N)^2$. In this process, an incoming photon scatters off the electromagnetic field of the tungsten nucleus and converts into an ALP through the $g_{a\gamma}$ vertex. This process proceeds through t-channel photon exchange, as
illustrated in Fig.~\ref{fig:feynman_gag}, and corresponds to the Primakoff production mechanism in the Coulomb field of nucleus. Since the production rate is enhanced by the atomic number square $Z^2$, a high-$Z$ material such as tungsten provides an efficient target.
The same production mechanism has been studied through the LUXE-NPOD optical dump setup in Ref.~\cite{Bai:2021gbm}.
The high-energy photon can also scatter off atomic electrons. However, the cross section for Primakoff scattering off electrons is significantly smaller than that for scattering off nucleus which is enhanced by a factor of $Z^2$. We will not consider Primakoff scattering off electrons below.

For the Primakoff process, the scattering amplitude can be written as
\begin{eqnarray}
i\mathcal M^{\rm Prim}_{\gamma N\to aN} = ie \frac{g_{a\gamma}}{t} J_{\rm had}^{\mu}J_{\gamma a~\mu}\;,
\end{eqnarray}
where $J_{\rm had}^{\mu}$ and $J_{\gamma a}^{\mu}$ denote the hadronic current and the photon-ALP current, respectively. After averaging over the initial photon polarization and the nuclear spin, the squared amplitude takes the tensor form
\begin{eqnarray}
\overline{|\mathcal M^{\rm Prim}_{\gamma N\to aN}|^2} = e^2 g_{a\gamma}^2\frac{1}{t^2} L_{\gamma a}^{\mu\nu}W_{\mu\nu}\;,
\end{eqnarray}
where $L_{\gamma a}^{\mu\nu}$ is fixed by the $a\gamma\gamma$ vertex, while $W_{\mu\nu}$ contains the elastic nuclear response and the nuclear charge form factor $F_N$. See Appendix~\ref{app:primakoff} for the details. Carrying out the tensor contraction gives the laboratory-frame differential cross section~\cite{Aloni:2019ruo}
\begin{eqnarray}
\frac{d\sigma^{\rm Prim}_{\gamma N\to aN}}{dt} = \alpha Z^2 |F_N(t)|^2 \Gamma(a\to\gamma\gamma) \mathcal H(m_N,m_a,s,t)\;,
\label{eq:dsigma_prim}
\end{eqnarray}
where
\begin{eqnarray}
\mathcal H(m_N,m_a,s,t) = \frac{128\pi m_N^4}{m_a^3}
\frac{m_a^2t(m_N^2+s)-m_a^4m_N^2 -t\left[(s-m_N^2)^2+st\right]}{t^2(s-m_N^2)^2(t-4m_N^2)^2}
\end{eqnarray}
with $s=m_N^2+2m_NE_\gamma$ for a stationary target nucleus. We integrate the product of the differential cross section and the decay probability over the variable $t$. The kinematic limits are
\begin{eqnarray}
t_{\rm min,max} = m_a^2 - \frac{s-m_N^2}{2s} \left[ s+m_a^2-m_N^2 \pm \lambda^{1/2}(s,m_a^2,m_N^2) \right],
\label{eq:tlimits}
\end{eqnarray}
with the K\"all\'en function $\lambda(x,y,z)=x^2+y^2+z^2-2xy-2xz-2yz$.

After substituting Eq.~(\ref{eq:dsigma_prim}) into the general event-yield formula in Eq.~(\ref{eqn:Na}), we evaluate the expected number of ALP events in the photon-philic ALP scenario.
In this case, the ALP production rate is determined by the UV parameter $g_{a\gamma}$ through the Primakoff cross section. The decay probability is given by the same coupling through the decay width $\Gamma(a\to\gamma\gamma)$.
For the SM backgrounds, we follow the estimate for LUXE-NPOD in Ref.~\cite{Bai:2021gbm}.
For a $1~{\rm m}$ tungsten dump, the primary Compton photons are expected to be absorbed before reaching the decay volume.
The SM backgrounds mainly come from three sources: charged particles produced in the shower, neutrons misidentified as
photons, and real photons from electromagnetic or hadronic interactions near the end of the dump.
Charged particles can be swept away by a magnetic field, so the dominant residual background is the neutral component.
In this setup, the rate of neutral background reaching the detector is estimated to be $\mu_n\simeq10$ for neutrons or $\mu_\gamma\simeq0.013$ for real photons per bunch crossing.
The neutron rate must be weighted by the probability that a neutron is misidentified as a photon.
For a neutron-to-photon fake rate $f_{n\to\gamma}\lesssim10^{-3}$, the effective fake-photon rate becomes $\mu_n f_{n\to\gamma}\lesssim10^{-2}$ per bunch crossing, which is the same order as the real-photon rate.
Both components are then further suppressed by the event-selection rejection factor $R_{\rm sel}\lesssim10^{-3}$.
Under these assumptions, the expected background is well below one event per year.
We therefore adopt the same background-free benchmark to get the sensitivity reach.

\begin{figure}[ht]
\begin{center}
\includegraphics[width=0.475\textwidth]{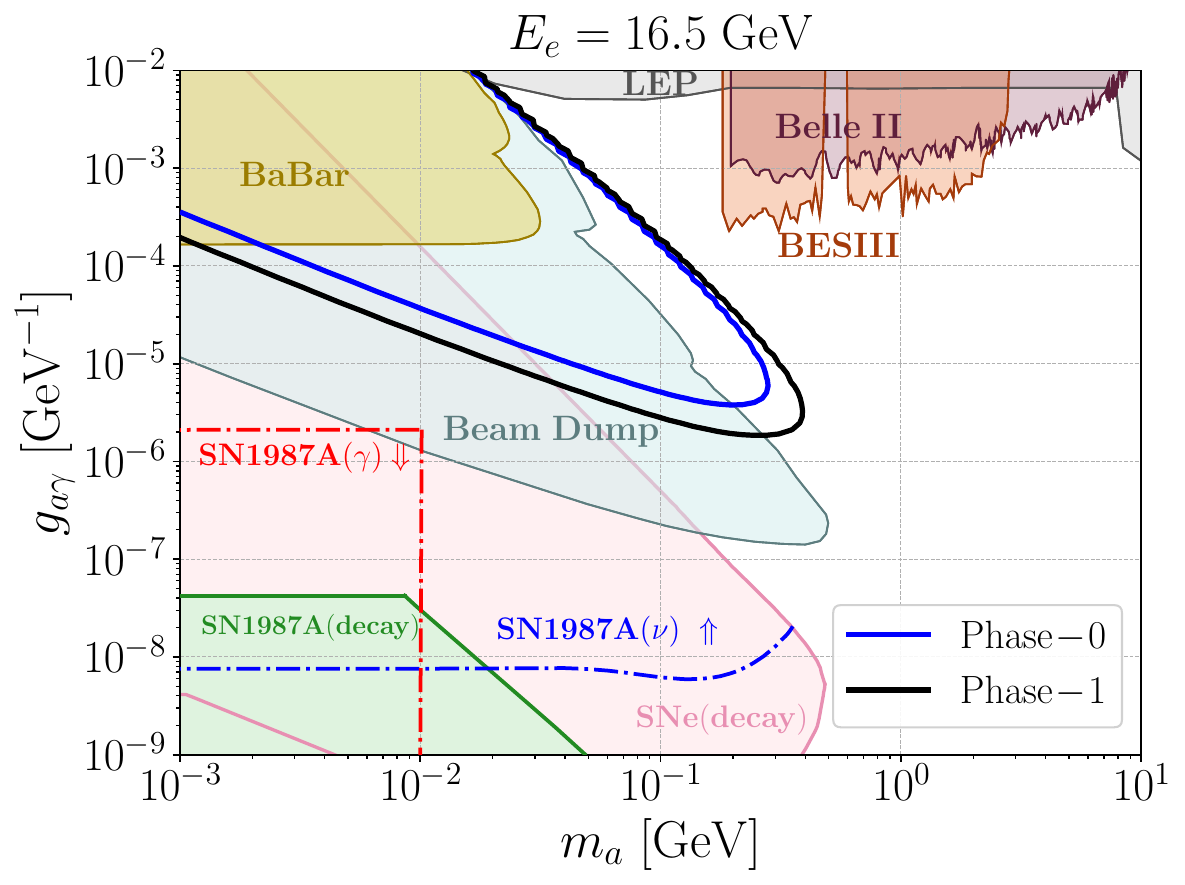}
\includegraphics[width=0.475\textwidth]{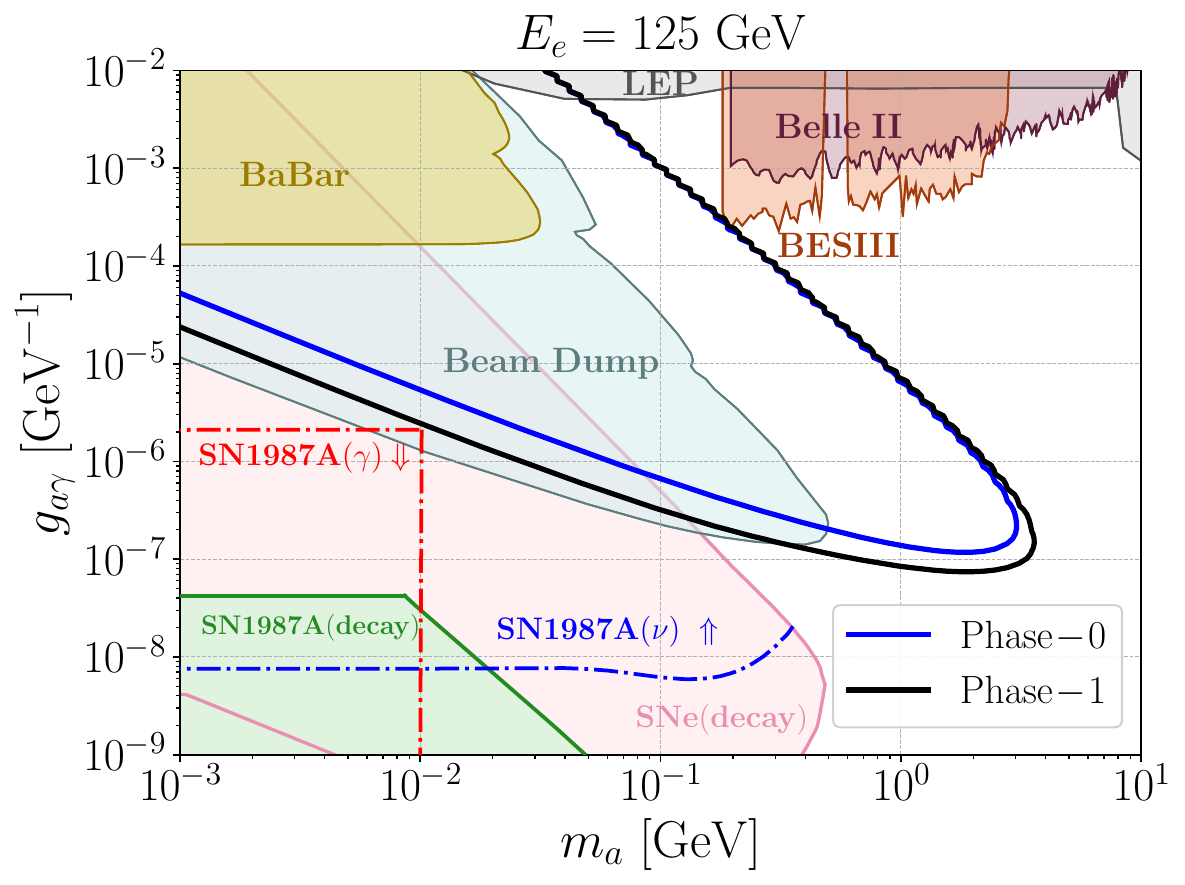}
\end{center}
\caption{
The sensitivity reach in the parameter space of $m_a$ versus $g_{a\gamma}$ from ALP production via the Primakoff process $\gamma + N\to a + N$ and $a\to \gamma\gamma$ decay mode, assuming only the ALP-photon coupling $g_{a\gamma}$. We take two different laser-electron collision configurations: phase-0 (blue) and phase-1 (black). For a required number of signal events $N_a = 3$, the results are shown for electron beam energies $E_e = 16.5$ GeV (left panel) and $125$ GeV (right panel). We also show the exclusion limits from LEP~\cite{Jaeckel:2015jla} (gray), BESIII~\cite{BESIII:2022rzz,BESIII:2024hdv} (orange), Belle II~\cite{Belle-II:2020jti} (purple), existing beam dump (cyan, combined limit from CHARM~\cite{CHARM:1985anb}, E141~\cite{Riordan:1987aw}, E137~\cite{Dolan:2017osp,Bjorken:1988as}, NuCal~\cite{Dobrich:2019dxc,Blumlein:1990ay}, NA64~\cite{NA64:2020qwq}. see also the summaries~\cite{Dobrich:2019dxc,dEnterria:2021ljz}), and BaBar~\cite{BaBar:2017tiz,Dolan:2017osp} (yellow) for comparison. Some astrophysical constraints are also shown, including SN1987A gamma bound~\cite{Caputo:2021rux} (red dash-dotted line), SN1987A neutrino bound~\cite{Caputo:2022mah,Lucente:2020whw} (blue dash-dotted line), SN1987A bound with $a\to \gamma\gamma$~\cite{Muller:2023vjm,Hoof:2022xbe} (green) and low-energy supernovae bound with $a\to \gamma\gamma$~\cite{Caputo:2022mah} (magenta).
}
\label{figs:ma_gag}
\end{figure}

We take the tungsten dump geometry and the photon spectra described in Sec.~\ref{sec:Dump}, and require three expected signal events $N_a=3$ to obtain the sensitivity contour.
The sensitivity reach in the plane of $m_a$ versus $g_{a\gamma}$ is shown in Fig.~\ref{figs:ma_gag}. We show the results for $E_e=16.5~{\rm GeV}$ (left panel) and for $E_e=125~{\rm GeV}$ (right panel). We find good agreement between our low-energy beam results and those reported in Ref.~\cite{Bai:2021gbm}.
The parameter space inside the contours yields more than three expected signal events. The ALPs with a coupling above
the upper bound decay before reaching the LLP detector. Any $g_{a\gamma}$ coupling smaller than the
lower bound $\sim 10^{-6}~{\rm GeV}^{-1}$ ($\sim 10^{-7}~{\rm GeV}^{-1}$) for $E_e=16.5~(125)$ GeV makes the production rate too small to generate enough signal events.
The phase-1 configuration reaches smaller couplings than phase-0, as expected from its larger hard-photon flux. The sensitivity to ALP mass is enhanced by one order of magnitude when switching to a higher-energy beam.

%%%%%%%%%%%%%%%%%%%%%%%%%%%%%%%
\subsection{Compton scattering}
%%%%%%%%%%%%%%%%%%%%%%%%%%%%%%%

The above scenario assumes that ALP production is dominated by the ALP-photon coupling.
If instead the ALP only couples to electrons in the electron-philic case, it can also be produced through a Compton-like scattering process
\begin{eqnarray}
\gamma+e\to a+e\;,
\end{eqnarray}
where the atomic electrons reside outside the tungsten nucleus.
Next, we examine this Compton scattering for the electron-philic ALP. Note that the effective ALP-photon coupling can also be induced by ALP-electron vertex at one-loop level. Here, for simplicity, we ignore the Primakoff process given by the loop-suppressed $g_{a\gamma}^{\rm eff}$ coupling and only take into account the effect on ALP decay. The correlation between not-negligible ALP-photon coupling and ALP-electron coupling will be discussed in the next subsection.

\begin{figure}[htb]
\centering
\includegraphics[width=0.48\textwidth]{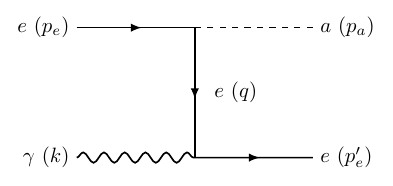}
\includegraphics[width=0.48\textwidth]{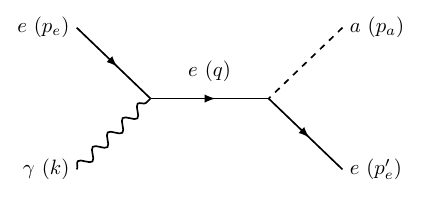}
\caption{The Feynman diagrams for ALP production via photon scattering off atomic electrons, $\gamma(k) + e^{-}(p_e) \to a(p_a)+e^{-}(p'_e)$.
}
\label{fig:FD2}
\end{figure}

Now, we present the analytical framework for estimating the expected signal yield of ALPs produced via the photon-electron scattering process $\gamma + e^- \to a + e^-$ in a tungsten target. We begin with considering the idealized case of scattering off a free electron, which provides the fundamental cross section for this process. The relevant Feynman diagrams are shown in Fig.~\ref{fig:FD2}. The differential cross section with respect to the Mandelstam variable $t$ for this free-electron Compton scattering is given as
\begin{eqnarray}
\frac{d\sigma^{\rm free}_{\gamma e\to a e}}{dt}
&=& \frac{g_{ae}^2\,\alpha}{2 (s - m_e^2)^2} \Bigg[\frac{m_a^2(3m_e^2 - s) + (m_e^2 - s)(m_e^2 - s - t)}{(m_e^2 + m_a^2 - s - t)^2} \nonumber\\
&& + \frac{m_a^2(3m_e^2 - s) + (m_e^2 - s)(m_e^2 - s - t)}{(s - m_e^2)^2}\nonumber\\
&&  + \frac{4m_a^2 m_e^2 + 2(m_a^2 + m_e^2 - s)(m_e^2 - s - t)}{(s - m_e^2)(m_e^2 + m_a^2 - s - t)} \Bigg]\;,
\end{eqnarray}
where $s=(k+p_e)^2$ and $t=(p_e-p_e')^2$.
While the free-electron approximation provides the fundamental interaction strength, a realistic fixed-target environment must account for the fact that electrons are bound within atomic shells. Consequently, a physically accurate cross section must incorporate atomic effects, such as nuclear screening and electron binding energies. For the high-energy photons considered in this work, on the order of $\mathcal{O}(1-10)$ GeV or even beyond, the incident energy far exceeds the K-shell binding energy of the tungsten target. Thus, the influence of individual atomic binding energies becomes secondary, which allows us to treat the collective response of the atom through a corrected cross section~\cite{Dent:2019ueq,Chakrabarty:2019kdd}
\begin{eqnarray}
{d\sigma_{\gamma e\to a e}^{\rm Comp}(E_{\gamma})\over dt} =
\frac{d\sigma^{\rm free}_{\gamma e\to a e}}{dt} \times Z_{\rm eff}\left( T_e  \right)  \,,
\label{eqn:sigma_gae}
\end{eqnarray}
where the integration limits for $t$ take the same form as those in Eq.~(\ref{eq:tlimits}), with the nuclear mass $m_N$ replaced by the electron mass $m_e$.
$Z_{\rm eff}$ is the effective atomic number.
While the fundamental interaction occurs with individual electrons, the collective effect of the atomic electrons is captured by the effective charge $Z_{\rm eff}$. It is a function of the electron recoil kinetic energy $T_e$. Since atomic electrons occupy discrete shells, scattering only occurs if the energy transfer $T_e$ exceeds the binding energy of a specific shell. Thus, $Z_{\rm eff}(T_e)$ acts as a step-wise weight factor that determines the number of active electrons available for ionization at a given energy transfer.
It is given by the form $Z_{\rm eff}(T_e)=\sum_i^{Z}\Theta(T_e-B_i)$ with $B_i$ being the binding energy of the $i$th electron~\cite{Chen:2016eab}.
The specific discrete values of $Z_{\rm eff}$ and their corresponding energy thresholds for tungsten ($Z=74$)~\cite{XRayDataBooklet:LBNL,NIST:XrayTrans} are summarized in Table~\ref{tab:Zeff}.
To evaluate the total cross section, we utilize the kinematic relationship between the electron recoil kinetic energy $T_e$ and the variable $t$
\begin{eqnarray}
t = (p_e - p_e')^2 = 2m_e^2 - 2m_e E_e' = 2m_e (m_e - E_e') = - 2m_e T_e\;.
\end{eqnarray}
By performing this variable substitution, we can express the effective charge $Z_{\rm eff}$ as a function of $t$. This allows us to integrate the differential cross section over the $t$ variable while accounting for the energy-dependent electron occupancy $Z_{\rm eff}$ in a single step.

\begin{table}[ht]
\centering
\begin{tabular}{c|c||c|c}
\hline
%\hline
Recoil Energy Range [eV] & $Z_{\text{eff}}$ & Recoil Energy Range [eV] & $Z_{\text{eff}}$ \\ \hline
$T_e > 69525$ & 74 & $490.4 < T_e \le 594.1$ & 28\\
$12100 < T_e \le 69525$ & 72 & $423.6 < T_e \le 490.4$ & 26\\
$11544 < T_e \le 12100$ & 70 & $255.9 < T_e \le 423.6$ & 20\\
$10207 < T_e \le 11544$ & 68 & $243.5 < T_e \le 255.9$ & 18\\
$2820 < T_e \le 10207$ & 60 & $75.6 < T_e \le 243.5$ & 10\\
$2575 < T_e \le 2820$ & 58 & $45.3 < T_e \le 75.6$ & 8 \\
$2281 < T_e \le 2575$ & 56 & $36.8 < T_e \le 45.3$ & 6 \\
$1872 < T_e \le 2281$ & 50 & $33.6 < T_e \le 36.8$ & 4 \\
$1809 < T_e \le 1872$ & 48 & $31.4 < T_e \le 33.6$ & 2 \\
$594.1 < T_e \le 1809$ & 30 & $T_e \le 31.4$ & 0 \\
%$490.4 < T_e \le 594.1$ & 28 \\
%$423.6 < T_e \le 490.4$ & 26 \\
%$255.9 < T_e \le 423.6$ & 20 \\
%$243.5 < T_e \le 255.9$ & 18 \\
%$75.6 < T_e \le 243.5$ & 10 \\
%$45.3 < T_e \le 75.6$ & 8 \\
%$36.8 < T_e \le 45.3$ & 6 \\
%$33.6 < T_e \le 36.8$ & 4 \\
%$31.4 < T_e \le 33.6$ & 2 \\
%$T_e \le 31.4$ & 0 \\
\hline
%\hline
\end{tabular}
\caption{Effective atomic number $Z_{\text{eff}}$ for tungsten ($Z=74$) as a function of the electron recoil energy $T_e$ in unit of eV.}
\label{tab:Zeff}
\end{table}

\begin{figure}[ht]
\begin{center}
\includegraphics[width=0.475\textwidth]{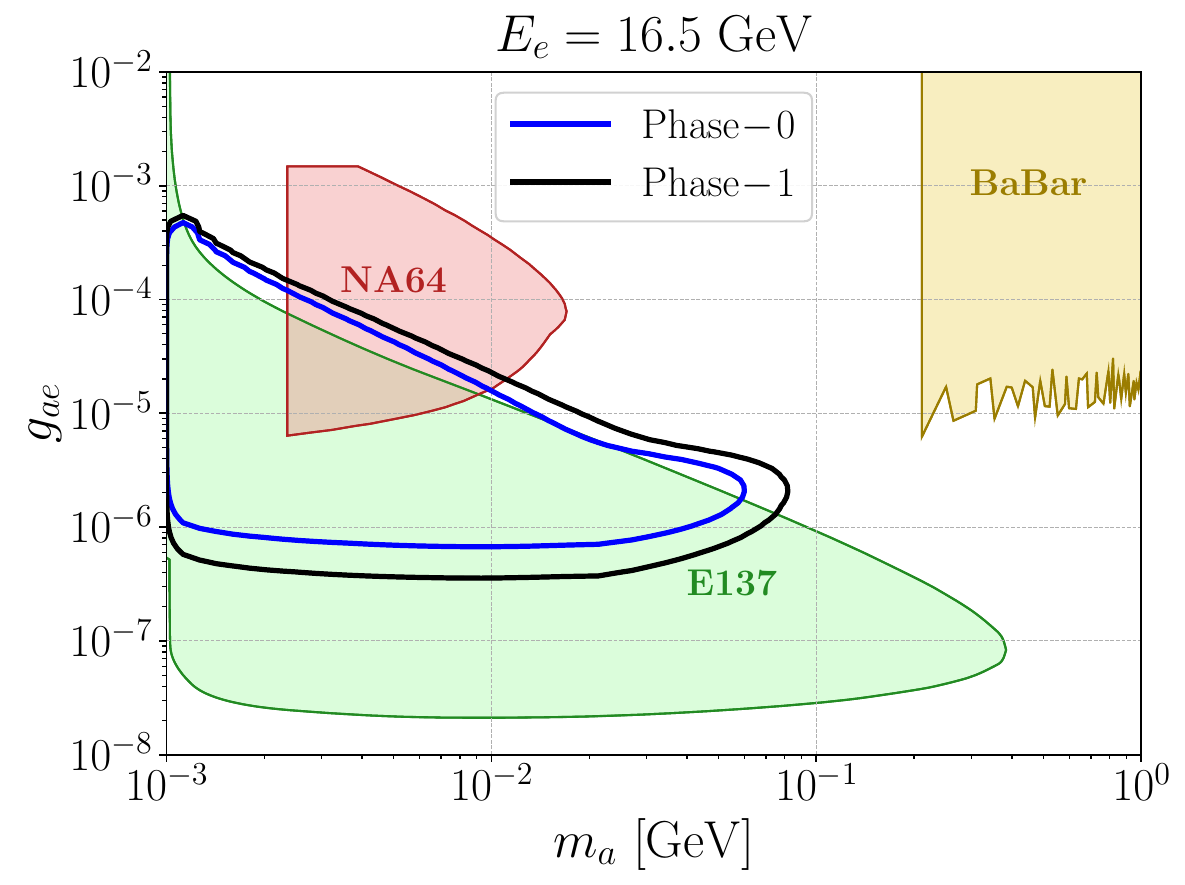}
\includegraphics[width=0.475\textwidth]{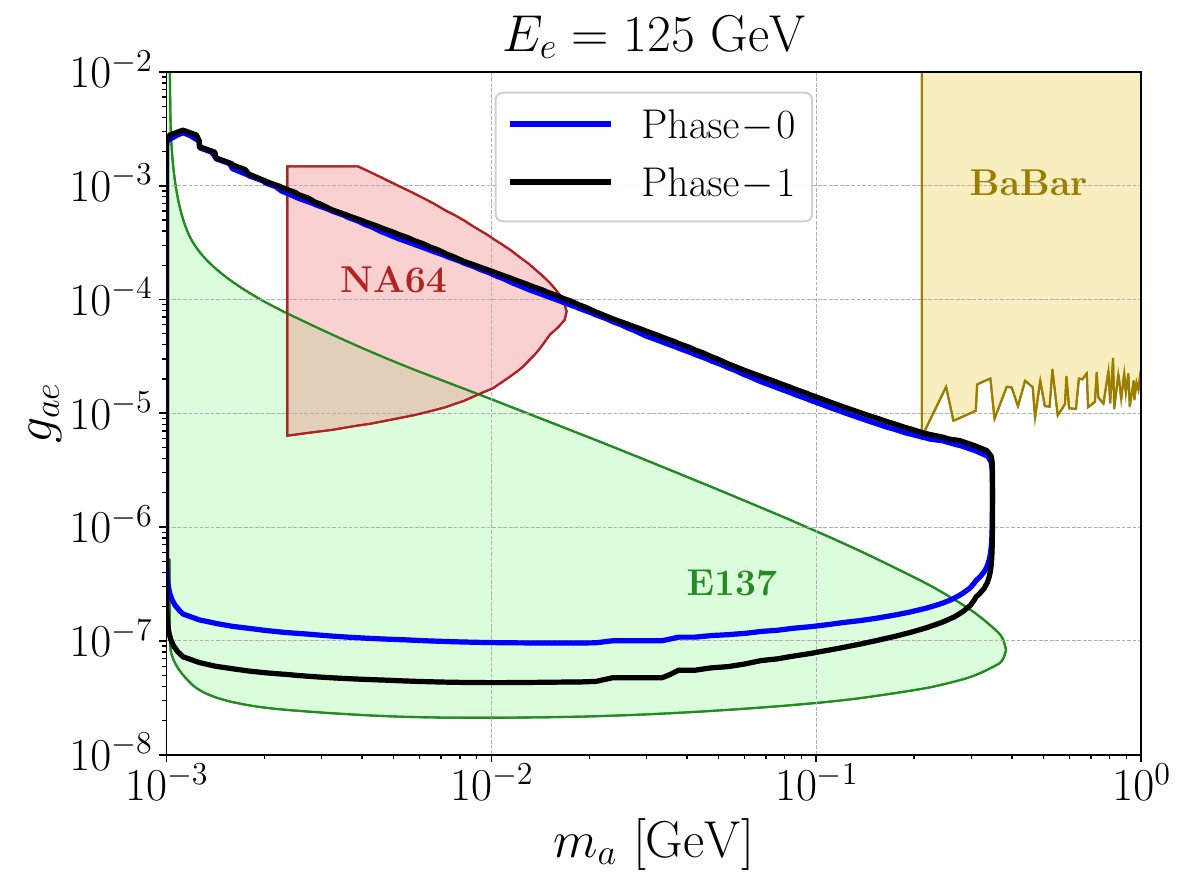}
\end{center}
\caption{
The sensitivity reach in the parameter space of $m_a$ versus $g_{ae}$ from ALP production via the Compton process $\gamma + e\to a + e$ and $a\to e^+e^-$ decay mode, assuming only the ALP-electron coupling $g_{ae}$. We take two different laser-electron collision configurations: phase-0 (blue) and phase-1 (black). For a required number of signal events $N_a = 3$, the results are shown for electron beam energies $E_e = 16.5$ GeV (left panel) and $125$ GeV (right panel). We also show the exclusion limits from BaBar (yellow)~\cite{Bauer:2017ris,Bauer:2018uxu,Liu:2023bby}, E137 (green)~\cite{Liu:2017htz}, and NA64 (red)~\cite{NA64:2021aiq} for comparison.
}
\label{fig:ma_gae}
\end{figure}

Based on the cross section for $\gamma +e^-\to a +e^-$, we compute the signal rate from Eq.~(\ref{eqn:Na}) by assuming only the ALP-electron coupling $g_{ae}$ and searching for $a\to e^+e^-$ decay mode.
The tungsten target is treated as a collection of atomic electrons, with the number of active electrons determined by the effective atomic number $Z_{\rm eff}(T_e)$.
We then extract the projected sensitivity by solving for the contour with $N_a=3$ expected events.
The results are displayed in Fig.~\ref{fig:ma_gae} in the plane of $m_a$ versus $g_{ae}$.
One can see that the optical dump setup can probe ALP-electron coupling down to $g_{ae}\sim 10^{-6}$ ($10^{-7}$) and even lower for $E_e=16.5~(125)$ GeV.
Compared with phase-0, the phase-1 extends to lower ALP-electron coupling because of its larger flux of GeV photons. The sensitivity to ALP mass is also enhanced by a factor of five when switching to a higher-energy beam.

%%%%%%%%%%%%%%%%%%%%%%%%%%%
\subsection{Correlation between Primakoff and Compton processes}
%%%%%%%%%%%%%%%%%%%%%%%%%%%

We now consider the case in which more than one ALP coupling is present at the same time.
First, we focus on the two-dimensional parameter space spanned by $g_{a\gamma}^{\rm eff}$ and $g_{ae}$.
In this case, ALPs can be produced through both Primakoff production on nucleus and Compton-like scattering on atomic electrons.
In our event-rate estimate, the total production rate is obtained by adding the two independent production contributions
\begin{eqnarray}
\sigma^{\rm Prim+Comp}(E_\gamma) = \sigma^{\rm Prim}_{\gamma N\to aN}(E_\gamma) + \sigma^{\rm Comp}_{\gamma e\to ae}(E_\gamma)\;,
\end{eqnarray}
where the first (second) term is the Primakoff (Compton-like) contribution determined by $g_{a\gamma}^{\rm eff}$ ($g_{ae}$) and is evaluated using Eq.~(\ref{eq:dsigma_prim}) (Eq.~(\ref{eqn:sigma_gae})).
The decay widths and branching fractions are calculated using the total ALP decay width, including both the $a\to\gamma\gamma$ and $a\to e^+e^-$ channels whenever they are kinematically allowed. Note that Ref.~\cite{Bauer:2021mvw} calculated the off-shell ALP-photon vertex function with off-shell ALP, one on-shell photon and one off-shell photon. According to their result, in the limit where the ALP is taken on-shell, one recovers the expression for the effective coefficient $c_{a\gamma}^{\rm eff}$ entering the $a\to \gamma\gamma$ decay rate. Thus, it is appropriate to use the effective coupling $g_{a\gamma}^{\rm eff}$ in the Primakoff production.

\begin{figure}[ht]
\begin{center}
\includegraphics[width=0.475\textwidth]{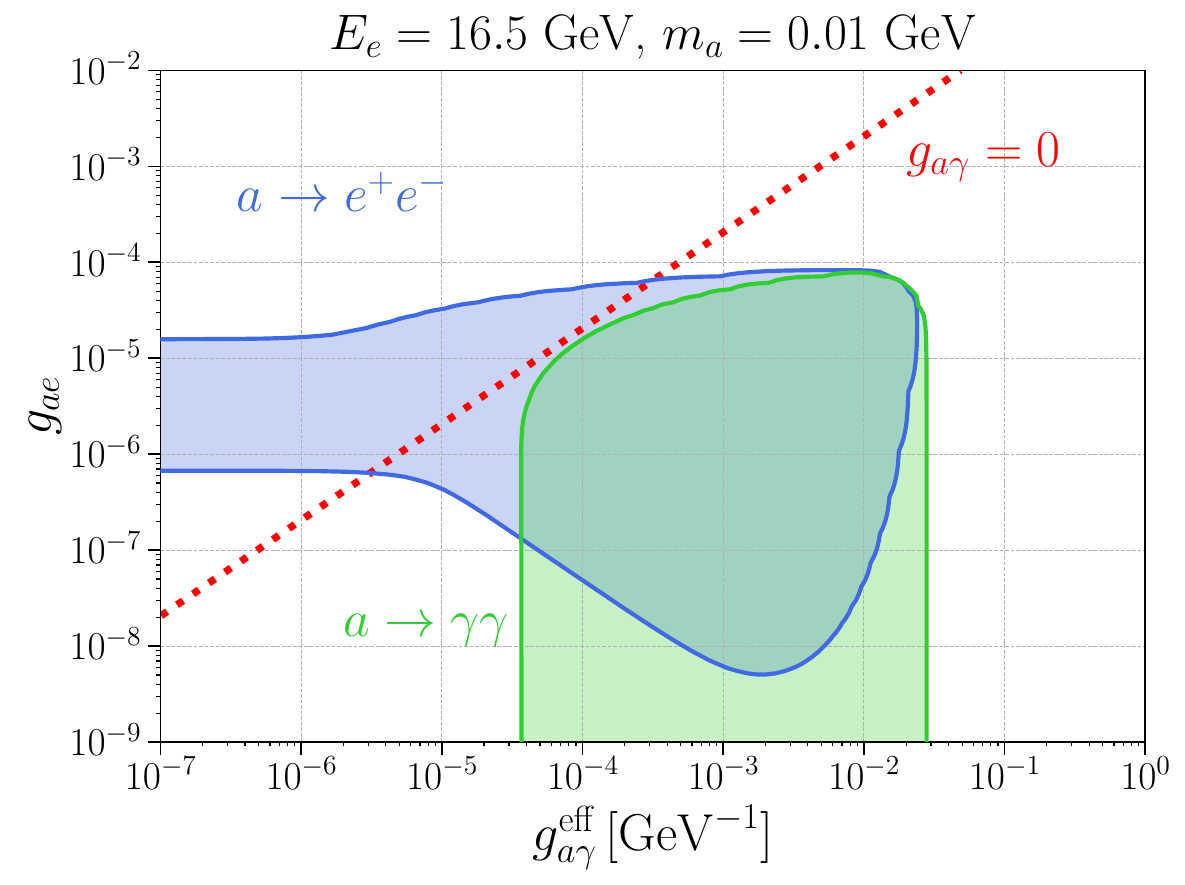}
\includegraphics[width=0.475\textwidth]{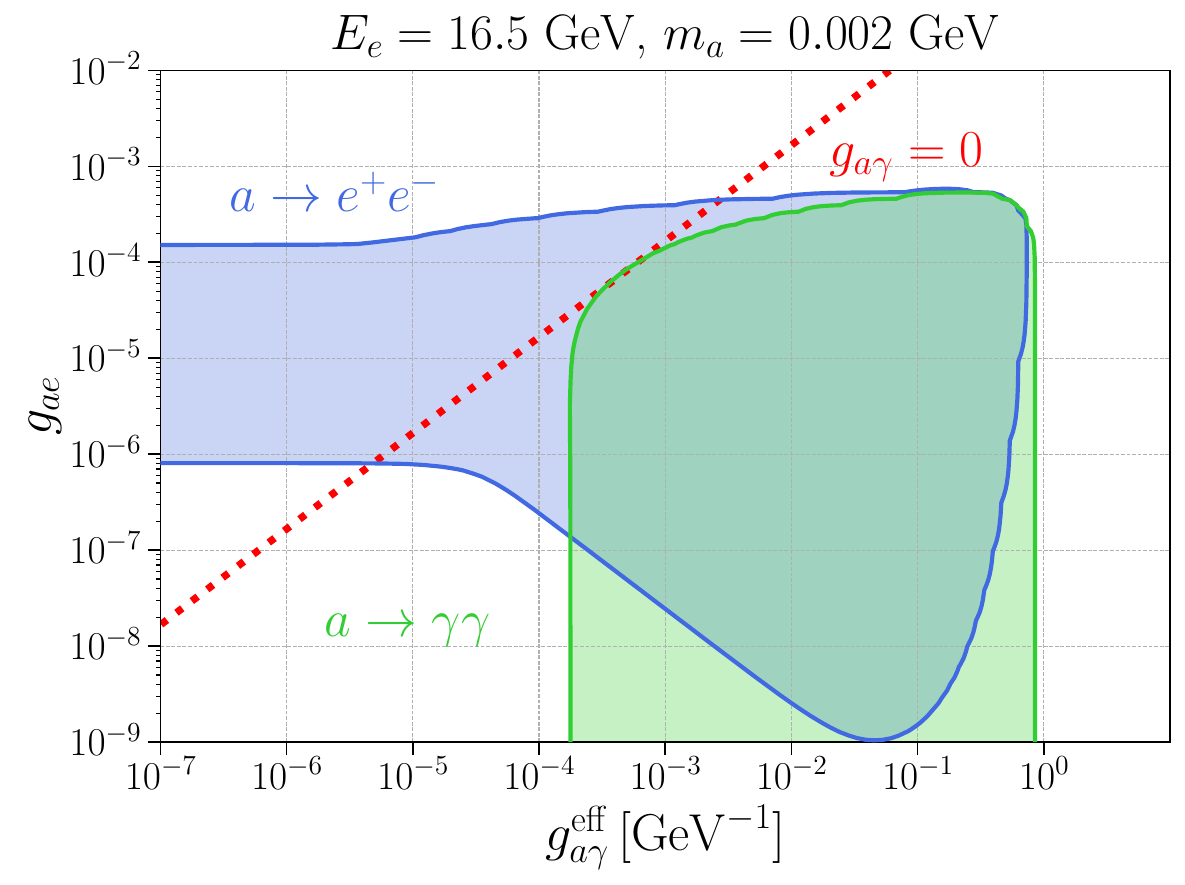}\\
\includegraphics[width=0.475\textwidth]{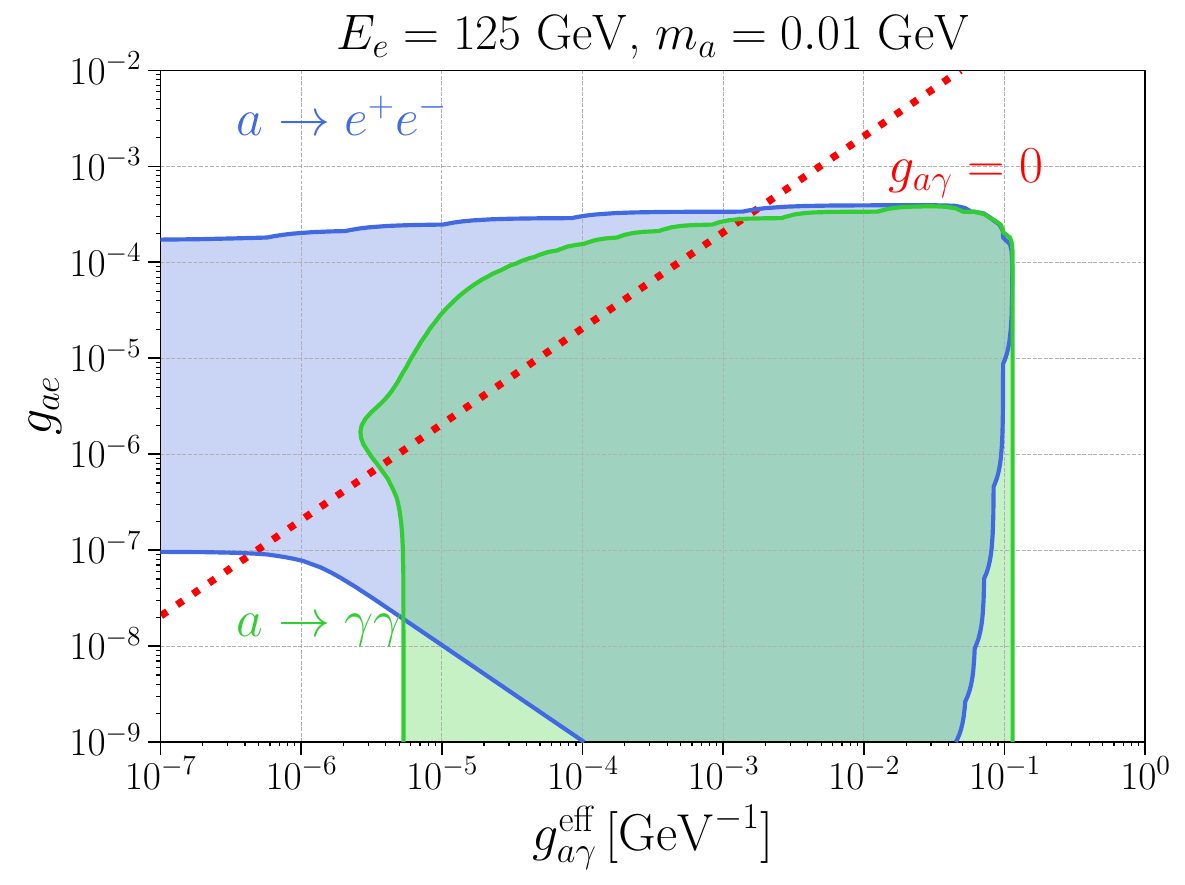}
\includegraphics[width=0.475\textwidth]{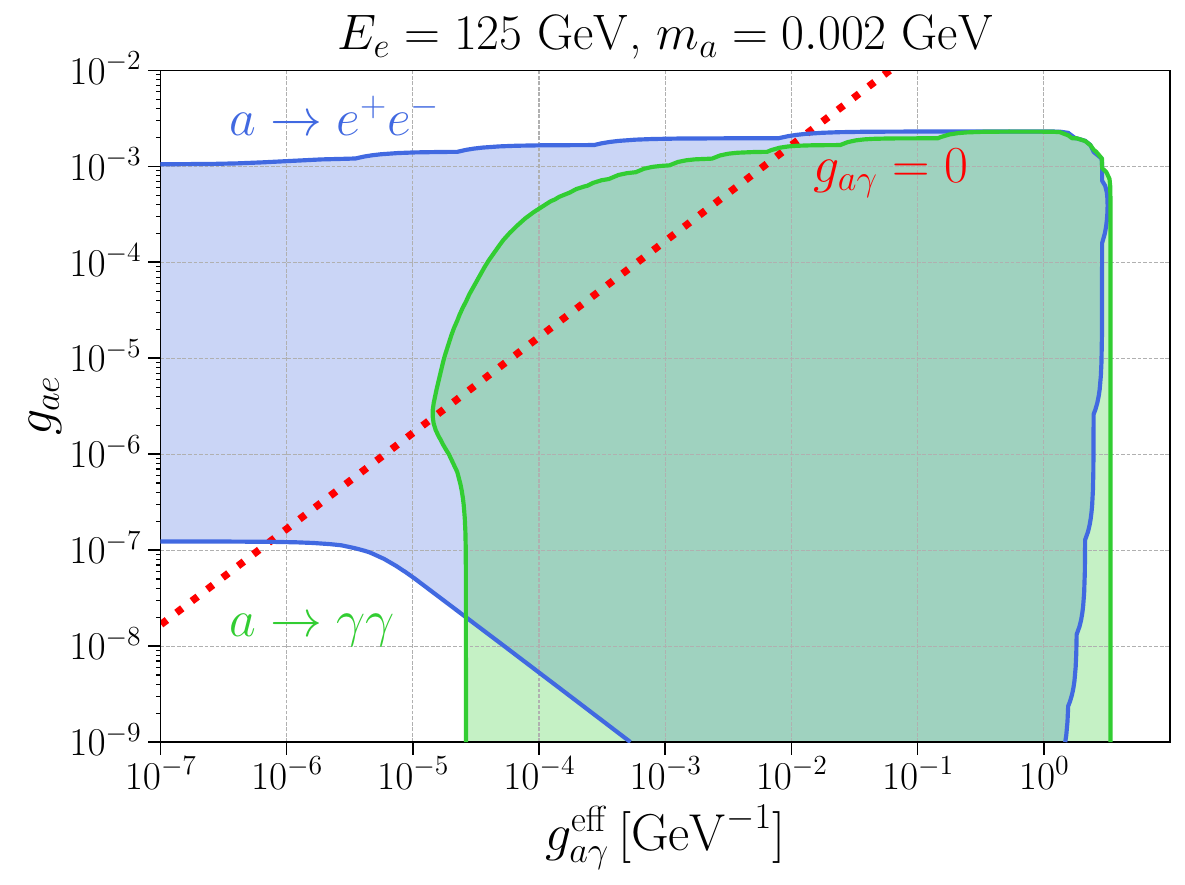}\\
\end{center}
\caption{
The sensitivity reach in the parameter space of $g_{a\gamma}^{\rm eff}$ versus $g_{ae}$ in phase-0, assuming electron beam energies $E_e = 16.5$ GeV (upper panels) and 125 GeV (lower panels).
The blue (green) shaded regions correspond to the decay channel of $a \to e^+ e^-$ ($a \to \gamma \gamma$).
The left (right) panels show results for the ALP mass benchmark $m_a = 0.01$ GeV ($0.002$ GeV). The red dotted lines indicate the special case with $g_{a\gamma}=0$ and $g_{a\gamma}^{\rm eff}$ completely given by $g_{ae}$.
}
\label{figs:gag_gae}
\end{figure}

The resulting sensitivity contours in the plane of $g_{a\gamma}^{\rm eff}$ versus $g_{ae}$ are shown in Fig.~\ref{figs:gag_gae}.
Instead of scanning the full three-dimensional parameter space of $m_a$, $g_{a\gamma}^{\rm eff}$ and $g_{ae}$, we choose two representative ALP mass benchmarks $m_a=0.01$ GeV (left panels) and $0.002$ GeV (right panels).
This allows us to display directly how the two couplings compensate each other in producing an observable signal.
The contours correspond to $N_a=3$ expected signal events in phase-0, for $E_e=16.5$ GeV (upper panels) and $E_e=125$ GeV (lower panels).
The contour shapes are determined by the combined effects of ALP production and decay. The blue and green shaded regions correspond to the decay channel of $a\to e^+e^-$ and $a\to \gamma\gamma$, respectively. We also show red dotted lines indicating the special case with $g_{a\gamma}=0$ and $g_{a\gamma}^{\rm eff}$ completely given by $g_{ae}$. The search for $a\to e^+e^-$ ($a\to \gamma\gamma$) decay channel places an upper limit on the sensitivity of $g_{a\gamma}^{\rm eff}$ ($g_{ae}$) coupling.
When $g_{ae}$ is very small, the Compton-like production channel become ineffective, and the sensitivity reduces to the single ALP-photon coupling case discussed above.
Conversely, when $g_{a\gamma}^{\rm eff}$ is very small, the Primakoff contribution becomes negligible, and the constraint approaches the limit of single ALP-electron coupling.
In the intermediate region, both couplings contribute to the event yield, and the sensitivity cannot be interpreted as a bound on only one coupling. Notably, due to the sizable contribution from Primakoff process, the sensitivity to the $g_{ae}$ coupling in the intermediate region can be improved by 2-3 orders of magnitude compared to the case where $g_{a\gamma}^{\rm eff}$ can be neglected.

%%%%%%%%%%%%%%%%%%%%%%%%%%%%%%%%
\section{Conclusion}
\label{sec:Con}
%%%%%%%%%%%%%%%%%%%%%%%%%%%%%%%%

The feeble interactions of light ALPs render them long-lived.
Probing long-lived ALPs therefore demands facilities with a macroscopic decay volume to match their extended decay lengths, such as high-intensity beam dump experiments.
Conventional beam dump experiments utilize high-energy fermion beams to collide on a fixed target.
ALP production proceeds predominantly via $2\to 3$ bremsstrahlung and Primakoff-like processes which require dedicated approximation method to simplify the calculation. An optical dump was proposed based on the collision of a high-energy electron beam and high-intensity laser pulse. The large flux of hard photons from the electron-laser collision was suggested to strike a dump of target atom tungsten to produce feebly interacting new particles such as ALPs. The long-lived ALPs coupled to photons can be produced through a direct Primakoff
process without virtual photon approximation in the calculation.

In this work, we revisit the probe of long-lived ALPs with MeV $\sim$ GeV mass via a laser-assisted optical dump. We consider the low-energy effective
Lagrangian for ALPs incorporating both the ALP-photon and ALP-fermion (e.g., electron) interactions. The primary photon spectrum is obtained by using Ptarmigan package to implement the strong-field QED simulation of nonlinear Compton photon emission at LUXE. We then calculate the visible ALP signal rate by convolving the primary photon spectrum with ALP production cross section, decay probability and the detector efficiency. The scope of optical dump searches is extended to both the $g_{a\gamma}$ coupling induced Primakoff process and the Compton-like scattering via the $g_{ae}$ coupling.
We also demonstrate the complementarity
of Primakoff process and Compton scattering, and exhibit the interplay of two ALP couplings in light of optical dump experiment. Our conclusions are summarized as follows
\begin{itemize}
\item For photon-philic ALPs, the ALP production rate is only determined by the UV parameter $g_{a\gamma}$ through the Primakoff process. We find good agreement between our sensitivity to the ALP-photon coupling with a low-energy electron beam and that reported in Ref.~\cite{Bai:2021gbm} for the LUXE experiment. The
lower bound of $g_{a\gamma}$ becomes $\sim 10^{-6}~{\rm GeV}^{-1}$ ($\sim 10^{-7}~{\rm GeV}^{-1}$) for $E_e=16.5~(125)$ GeV. The sensitivity to ALP mass is enhanced by one order of magnitude when switching to a high-energy beam $E_e=125$ GeV.
\item For the electron-philic case, the ALPs can be produced through a Compton-like scattering process between the hard photons and the electrons in the target atoms. The
optical dump setup can probe ALP-electron coupling down to $g_{ae}\sim 10^{-6}$ ($10^{-7}$) and even lower for $E_e=16.5~(125)$ GeV.
\item For the case with both ALP-photon and ALP-electron couplings, the Primakoff process and Compton scattering compensate each other in producing an observable signal. Due
to the sizable contribution from Primakoff process, the sensitivity to the $g_{ae}$ coupling can be improved by 2-3 orders of magnitude compared to the case where the ALP-photon coupling can be neglected.
\end{itemize}

%###################################################################
%%%%%%%%%%%%%%%%%%%%%%%%
\acknowledgments
%%%%%%%%%%%%%%%%%%%%%%%%

T.~L. is supported by the National Natural Science Foundation of China under Grant No. 12375096.

%%%%%%%%%%%%%%%%%%%%%%%%%%%%%%%%%%%%%%%%%%%%%%
\appendix
%%%%%%%%%%%%%%%%%%%%%%%%%%%%%%%%%%%%%%%%%%%%%%

%%%%%%%%%%%%%%%%%%%%%%%%%
\section{The amplitude of Primakoff process}
\label{app:primakoff}
%%%%%%%%%%%%%%%%%%%%%%%%%

In this appendix we follow Ref.~\cite{Aloni:2019ruo} to present the derivation of the Primakoff production amplitude. The amplitude of Primakoff production is given by
\begin{eqnarray}
    i\mathcal M^{\rm Prim}_{\gamma N\to aN} = ie \frac{g_{a\gamma}}{t} J_{\rm had}^{\mu}J_{\gamma a~\mu}\;,
\end{eqnarray}
where $J_{\rm had}^{\mu}$ is the electromagnetic current of the target nucleus and $J_{\gamma a}^{\mu}$ is the photon-ALP current. From the $a\gamma\gamma$ vertex, the photon-ALP current is
\begin{eqnarray}
    J_{\gamma a}^{\mu} = i\epsilon^{\mu\alpha\rho\sigma} \varepsilon_{\alpha}(k)q_{\rho}k_{\sigma}\;.
\end{eqnarray}
After averaging over the initial photon polarization and the nuclear spin, the squared amplitude takes the tensor form
\begin{eqnarray}
    \overline{|\mathcal M^{\rm Prim}_{\gamma N\to aN}|^2} = e^2 g_{a\gamma}^2\frac{1}{t^2} L_{\gamma a}^{\mu\nu}W_{\mu\nu}\;,
\end{eqnarray}
where
\begin{eqnarray}
    L_{\gamma a}^{\mu\nu}= \frac{1}{2}\sum_{\lambda} J_{\gamma a}^{\mu}J_{\gamma a}^{\nu *}, \qquad
    W_{\mu\nu}=\sum_{\rm had,avg} J_{{\rm had},\mu}^{*}J_{{\rm had},\nu}\,.
\end{eqnarray}
Here the factor $1/2$ accounts for the average over the two initial photon polarizations.
Evaluating the polarization sum gives
\begin{eqnarray}
    L_{\gamma a}^{\mu\nu}=-\left[\frac{(t-m_a^2)^2}{8}g^{\mu\nu}+\frac{t-m_a^2}{4} \left(p_a^\mu k^\nu+k^\mu p_a^\nu\right) +\frac{m_a^2}{2}k^\mu k^\nu \right]\,.
\end{eqnarray}
For coherent elastic scattering off the nucleus, the nuclear tensor is given by
\begin{eqnarray}
    W^{\mu\nu}\simeq 4\rho^2 Z^2 |F_N(t)|^2 K^\mu K^\nu,\qquad K^\mu=p_N^\mu+\frac{q^\mu}{2}\,,
\end{eqnarray}
with
\begin{eqnarray}
    \rho=\frac{16m_N^4}{(4m_N^2-t)^2}\,.
\end{eqnarray}
Substituting these expressions into differential cross section
yields the result Eq.~(\ref{eq:dsigma_prim}) given in the main text.
The finite size of the target nucleus is included through the nuclear charge form factor $F_N(t)$.
Following Ref.~\cite{Ness:2025klj}, we have
\begin{eqnarray}
|F_N(t)|^2 = \frac{1}{Z^2} \left[ G_2^{\rm el}(t)+G_2^{\rm inel}(t) \right]\,,
\end{eqnarray}
where
\begin{eqnarray}
G_2^{\rm el}(t) &=& \left(\frac{a^2 t}{1+a^2 t}\right)^2
\left(\frac{1}{1+t/d}\right)^2 Z^2\,, \nonumber \\
G_2^{\rm inel}(t) &=& \left(\frac{a'^2 t}{1+a'^2 t}\right)^2
\left[
\frac{1+\frac{t}{4m_p^2}(\mu_p^2-1)}
{\left(1+\frac{t}{0.71~{\rm GeV}^2}\right)^4}
\right]^2 Z \,.
\end{eqnarray}
The parameters are
\begin{eqnarray}
a &=& \frac{111 Z^{-1/3}}{m_e}\,,\nonumber \\
a'&=&\frac{773 Z^{-2/3}}{m_e}\,, \nonumber \\
d &=& 0.164~{\rm GeV}^2 A^{-2/3}\,,\nonumber \\
\mu_p &=& 2.79 \,,
\end{eqnarray}
where $Z=74$ and $A=184$ are taken for the tungsten target.
Here $F_N(t)$ is the nuclear electromagnetic charge form factor, which encodes the loss of coherence at finite momentum transfer. In the Primakoff amplitude, it always enters multiplied by the coherent enhancement factor $Z^2$.

%%%%%%%%%%%%%%%%%%%%%%%%%%%%%%%%%%%%%%%%%%%%%%
%%%%%%%%%%%%%%%%%%%%%%%%%%%%%%%%%%%%%%%%%%%%%%
%%%%%%%%%%%%%%%%%%%%%%%%%%%%%%%%%%%%%
\bibliography{refs}

\end{document}